\documentclass[twocolumn,10pt,amsmath,superscriptaddress,amssymb,longbibliography]{revtex4-2}
\usepackage{epsfig}
\usepackage{epstopdf}
\usepackage{dcolumn}
\usepackage{amssymb}
\usepackage{bm}
\usepackage{amsmath,amssymb,amsfonts}
\usepackage{appendix}
\usepackage{graphicx}
\usepackage{algorithm}
\usepackage{enumerate}
\usepackage{makeidx}
\usepackage{braket}

\usepackage{titlesec}
\titlespacing*{\subsection}{0pt}{1\baselineskip}{0.5\baselineskip}

\usepackage{color,CJK}
\usepackage[usenames,dvipsnames]{xcolor}
\usepackage[driverfallback=dvipdfmx,colorlinks,linkcolor=blue,citecolor=blue,urlcolor=blue]{hyperref}

\begin{document}

\title{Tetratomic states of microwave dressed and associated ultracold $^{23}$Na$^{40}$K molecules}

\author{Zhengyu Gu}
\author{Xuansheng Zhou}
\author{Wei Chen}
\affiliation{State Key Laboratory of Quantum Optics Technologies and Devices, \\ Institute of Opto-electronics, Collaborative Innovation Center of Extreme Optics, Shanxi University, Taiyuan, Shanxi 030006, People's Republic of China}

\author{Wei Han}
\affiliation{State Key Laboratory of Quantum Optics Technologies and Devices, \\ Institute of Opto-electronics, Collaborative Innovation Center of Extreme Optics, Shanxi University, Taiyuan, Shanxi 030006, People's Republic of China}
\affiliation{Hefei National Laboratory, Hefei, China.}

\author{Fulin Deng}
\author{Tao Shi}
\affiliation{CAS Key Laboratory of Theoretical Physics, Institute of Theoretical Physics, Chinese Academy of Sciences, Beijing 100190, People's Republic of China }

\author{Pengjun Wang}
\email[Corresponding author email: ]{pengjun\underline{ }wang@sxu.edu.cn }
\affiliation{State Key Laboratory of Quantum Optics Technologies and Devices, \\ Institute of Opto-electronics, Collaborative Innovation Center of Extreme Optics, Shanxi University, Taiyuan, Shanxi 030006, People's Republic of China}
\affiliation{Hefei National Laboratory, Hefei, China.}

\author{Jing Zhang}
\email[Corresponding author email: ]{jzhang74@sxu.edu.cn}
\affiliation{State Key Laboratory of Quantum Optics Technologies and Devices, \\ Institute of Opto-electronics, Collaborative Innovation Center of Extreme Optics, Shanxi University, Taiyuan, Shanxi 030006, People's Republic of China}
\affiliation{Hefei National Laboratory, Hefei, China.}

\date{\today }

\begin{abstract}

Ultracold diatomic molecules have achieved significant breakthroughs in recent years, enabling the exploration of quantum chemistry~\cite{Krems2008,liu2022Annu.Rev.Phys.Chem.,karman2024Nat.Phys.a}, precision measurements~\cite{PhysRevLett.100.043202,ACME2014,ye2024Phys.Rev.Lett.,demille2024Nat.Phys.}, and strongly correlated many-body physics~\cite{moses2017NaturePhys,PhysRevA.102.023320,Cornish2024}. Extending ultracold molecular complexity to polyatomic molecules, such as triatomic and tetratomic molecules, has attracted considerable interest~\cite{Balakrishnan2016,safronova2018Rev.Mod.Phys.,gacesa2021Phys.Rev.Research,Doyle2022}. However, the realization of ultracold polyatomic molecules remains technically challenging due to their complex energy-level structures. While only a few experiments have successfully demonstrated the formation of polyatomic molecules by magnetoassociation~\cite{yang2022Science} or electroassociation~\cite{chen2024Nature}, here we present the first step toward producing tetratomic molecules through the development of a microwave association technique combined with microwave dressing~\cite{uchlerPhysRevLett.98.2007,CooperPhysRevLett.103.2009,LevinsenPhysRevA.84.2011,yanPhysRevLett.125.2020}. When the two lowest rotational states of the molecules are dressed by a microwave field, weakly bound tetramer states emerge in the entrance channel with free dark excited states $\ket{0}$ and a dressed state $\ket{+}$. The spectroscopy of these weakly bound tetramers is probed by another microwave field that drives transitions from the populated dressed states $\ket{+}$. By precisely discriminating the complex hyperfine structure of the dark excited level $\ket{0}$ from the dressed-state spectroscopy, the binding energy of the tetratomic molecules is measured and characterized. Our work contributes to the understanding of complex few-body physics within a system of microwave-dressed molecules and may open an avenue toward the creation and control of ultracold polyatomic molecules.

\end{abstract}

\pacs{34.20.Cf, 67.85.Hj, 03.75.Lm}

\maketitle

Ultracold polar diatomic molecules serve as a versatile quantum resource with rich internal structure and large intrinsic electric dipole moments for a wide range of quantum-science applications, ranging from quantum simulation and quantum many-body physics to quantum chemistry. Extended to more complex polyatomic molecules, there are unexplored research frontiers in many-body physics and quantum chemistry~\cite{hutzler2020QuantumSci.Technol.,vilas2024Nature}. For example, ultracold polyatomic molecules offer an ideal platform for studying the quantum-mechanical few-body problem. Although calculating the potential energy surfaces for three-body, four-body, and five-body systems with high accuracy is extremely challenging~\cite{christianen2019J.Chem.Phys.,Unke2020,Wheatley2023,SardarPhysRevA.107.2023}, the experimental creation of ultracold polyatomic molecules presents a unique opportunity to explore the rich landscape of these complex surfaces.

However, the preparation and control of naturally occurring polyatomic molecules at ultracold temperature is extremely difficult because of the complexity of internal structure and collisional properties~\cite{PhysRevResearch.2.013384,PhysRevX.10.031022}. Recently, laser cooling and optoelctrical cooling methods naturally occurring polyatomic molecules have been successfully used, which has cooled various diatomic and linear triatomic molecules into the submillikelvin regime~\cite{prehn2016Phys.Rev.Lett.,KozyryevPhysRevLett.118.2017,Augenbraun2020NJP,vilas2022Nature,mitra2020Science,LangenNP2024}. Besides directly cooling naturally occurring polyatomic molecules, ultracold polyatomic molecules can also be realized by associating ultracold atoms or diatomic molecules in nanokelvin regime.

Triatomic molecules are created by magnetoassociation~\cite{yang2022Science}, which is achieved by adiabatically ramping the magnetic field through a Feshbach resonance between $^{23}$Na$^{40}$K molecules and $^{40}$K atoms~\cite{yang2019Science}. Tetratomic molecules are formed by electroassociation from diatomic $^{23}$Na$^{40}$K molecules in their rovibronic ground state~\cite{chen2024Nature}. By ramping the ellipticity of the microwave filed across the field-linked scattering resonance~\cite{chen2023Nature}, they transferred a pair of polar diatomic molecules in the dressed state $\ket{+}$ to the tetratomic molecules in the entrance channel $\ket{++}$. It is required a high microwave power to build the field-linked scattering resonance.  Nevertheless, it is important to develop new association techniques to prepare ultracold polyatomic molecules in a variety of distinct internal states.

Here we report microwave spectroscopy of associated tetratomic molecules from microwave-dressed fermionic $^{23}$Na$^{40}$K molecules, in which a microwave pulse couple two colliding dipolar molecules prepared in the dressed state $\ket{+}$ to a weakly bound tetramer state in the entrance channel $\ket{+0}$ with a dark state $\ket{0}$ and dressed state $\ket{+}$. The presence of many hyperfine states within dark states results in the existence of a variety of entrance channels, which give rise to the possibility of forming tetramer bound states. Compared to the field-linked scattering resonance method with single entrance channel, our approach operates efficiently without the need for high microwave power and shows insensitivity to the polarization of the microwave dressing field.

\begin{figure}[t]
\centerline{\includegraphics[width=3.1 in]{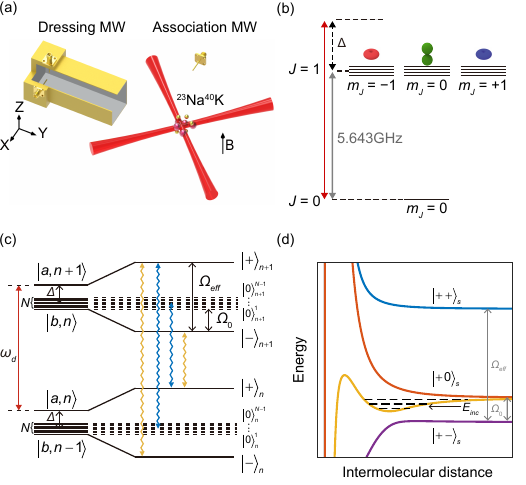}}
\caption{(Color online).
\textbf{Microwave dressing of $^{23}$Na$^{40}$K molecules.}
(a) Illustration of the experimental setup with microwave dressed molecules trapped in a crossed optical dipole trap at 1064 nm in the $x-y$ plane. Dressing MW denotes the microwave dressing field, which is generated by a doubly-fed waveguide propagates along the $y$ axis, while the induced molecular dipole is rotating in $x-z$ plane. Assocaition MW denotes the second microwave field, linearly polarized along the $\hat{x}$ direction, which is employed for the association of tetratomic molecules. The external magnetic field of 72.5 G is in the $z$ direction.
(b) The ground state $J=0$  and the first rotational state $J=1$ are separated by 5.643 GHz. The microwave dressing field with frequency $\omega_{d}$ and blue detuning $\Delta$ couples the $\ket{J=0,m_{J}=0}$ and $\ket{J=1,m_{J}=\pm1,0}$, where $m_{J}$ is rotational magnetic quantum number.
(c) Illustration of the allowed transitions in the dressed molecular system under a blue-detuned dressing microwave field. The left side depicts the uncoupled basis states, consisting of the ground state $\ket{a,n}$ and $N$ hyperfine excited states $\ket{b,n-1}$, where $n$ denotes the microwave photon number. The right side shows the corresponding dressed-state picture, including two bright dressed states $\ket{-}_{n}$ and $\ket{+}_{n}$, and $N{-}1$ dark states labeled as $\ket{0}^{1,\dots,N{-}1}_{n}$. The two dressed states are split by an energy interval of $\Omega_{\text{eff}}$, while the energy difference between the lower dressed state $\ket{-}_{n}$ and the dark state $\ket{0}^{1}_{n}$ is denoted by $\Omega_{0}$. Wavy lines represent the allowed transitions driven by the association microwave field: yellow lines indicate transitions between the two dressed states, and blue lines represent transitions between the dressed states and the dark states.
(d) Formation of tetramer states in $\ket{+0}_{s}$ channel. The solid lines denote the adiabatic potentials of the symmetric states $|ij\rangle _{\mathrm{s}}=(|ij\rangle+|ji\rangle )/\sqrt{2}$, and dashed lines denote a series of binding energies of the corresponding tetramer states.
\label{Fig1} }
\end{figure}

The experiment starts with fermionic molecules $^{23}$Na$^{40}$K in the their rovibronic ground state $\ket{J=0,m_{J}=0,m_{I_{\text{Na}}}=3/2,m_{I_{\text{K}}}=-4}$ about 2.2$\times10^{4}$ molecules at a temperature of 247 nK~\cite{li2023Sci.ChinaPhys.Mech.Astron.}, where $J$ is the total angular momentum excluding nuclear spin, $m _{J}$ is its projection onto the quantization axis set by external magnetic field, and $m_{I_{\text{Na}}}$ and $m_{I_{\text{K}}}$ are the projections of nuclear spins of $^{23}$Na and $^{40}$K. The energy structure of ground state and first rotational excited state is shown in Fig.~\ref{Fig1}(b) at a external magnetic field of 72.5 G. The rotational transition frequency between $J$=0 and $J$=1 is about $\omega_{0} = 2\pi \times 5.643 ~\mathrm{GHz}$~\cite{will2016Phys.Rev.Lett.}. Considering the coupling of rotation motion and nuclear spins, there are $N$=108 hyperfine states in the first rotational state $\ket{J=1}$ and 22 allowed transitions from the lowest ground state $\ket{m_{I_{\text{Na}}}=3/2,m_{I_{\text{K}}}=-4}$ considering all polarization case of microwave field, according to the transition selection rule $\Delta m_{F}=0,\pm1$. Here, $m_{F}=m_{J}+m_{I_{\text{Na}}}+m_{I_{\text{K}}}$ is the projection of total angular momentum onto the quantization axis. In the experiment, we identified the characteristic loss peaks within the spectrum of the rotational transition from the lowest hyperfine state $\ket{m_{I_{\text{Na}}}=3/2,m_{I_{\text{K}}}=-4}$ by adjusting the polarization of microwave field. From the microwave spectrum, we can identify the hyperfine states of the $J$=1 rotational state involved in the microwave dressing.

A microwave dressing field at frequency $\omega_{d}$  is applied to prepare the molecules into the dressed state $\ket{+}$ with blue detuning $\Delta=\omega_{d}-\omega_{0}$, which mixes the ground state and the first rotational state with different microwave photon numbers and induces a effective electrical dipole moment $d_{\mathrm{eff}}$. When the hyperfine coupling in the first rotational state is taken into account, this interaction results in the formation of two dressed states $\ket{+}$ and $\ket{-}$ alongside a dark state manifold with $N-1$ hyperfine states. The resulting dressed states can be regarded as linear superpositions of the ground and excited states. The $N-1$ dark states, which do not contain any ground-state component, can be divided into two categories: those that remain uncoupled from the microwave field, and those that are coherently mixed through microwave coupling.

\begin{figure}[htbp]
\centerline{\includegraphics[width=3.3 in]{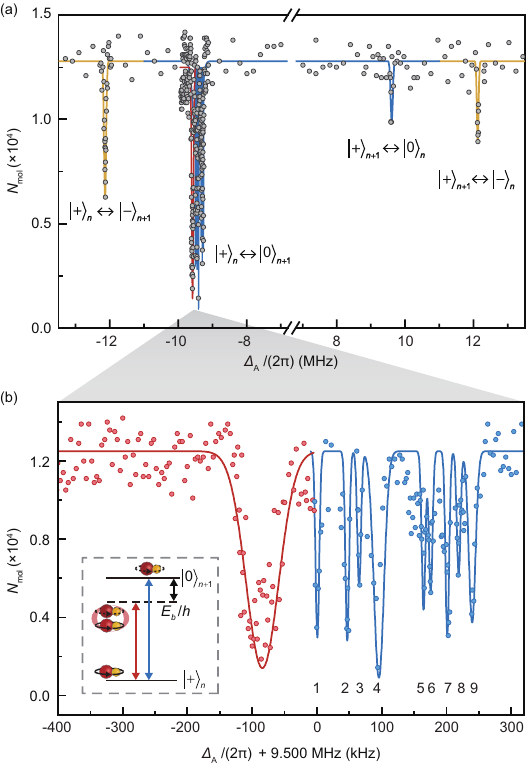}}
\caption{(Color online). \textbf{Dressed state spectroscopy and association of tetratomic molecules.}
(a) Loss spectrum of the molecules initially prepared in the dressed state $\ket{+}$ as a function of the microwave difference $\Delta_A = \omega_a - \omega_d$. The loss peaks indicate the resonant transition frequencies corresponding to the four allowed transitions from the dressed state $\ket{+}$. The microwave dressing field is applied with a coupling strength of $\Omega = 2\pi \times 9.913\,\mathrm{MHz}$ and a blue detuning of $\Delta = 2\pi \times 7\,\mathrm{MHz}$, while the association microwave field is applied with a duration of 500\,$\mu$s.
(b) Zoomed-in structure of the spectrum around the resonant transition $\ket{+}_{n}$ $\leftrightarrow$ $\ket{0}_{n+1}$. Besides the nine loss peaks marked by blue lines, which originate from transitions between dressed state and dark states, a new loss peak marked by a red line signifies the association of tetratomic molecules. The inset shows a simplified energy diagram illustrating the association of tetratomic molecules. The upper level corresponds to $\ket{0}_{n+1}$, and the lower level corresponds to $\ket{+}_{n}$. The bound tetramer state is represented by the dashed line. A microwave field can drive the association of tetratomic molecules at a frequency lower than the free-free transition $E_b/h$.
\label{Fig2} }
\end{figure}

In experiment, a microwave dressing field with adjusted polarization is generated by a doubly fed waveguide antenna, the relative phase and amplitude between two paths are then fine-tuned to achieve the arbitrary polarization, and propagates in the $\hat{y}$ direction as illustrated in Fig.~\ref{Fig1}(a). To study the dressed state spectroscopy, a second microwave field, denoted as the association microwave field, is linearly polarized along the $\hat{x}$ direction and applied to the molecules as a square pulse of 500 $\mu$s duration. At the end of the sequence, the dressing microwave field is ramped down and the residual molecules in dressed state $\ket{+}$ is adiabatically transferred to the uncoupled state $\ket{0, 0}$.
We record the number of remaining molecules in the state $\ket{0, 0}$ as a function of the microwave frequency difference $\Delta_{A}=\omega_{a}-\omega_{d}$, $\omega_{a}$ and $\omega_{d}$ are the frequencies of the association microwave field and microwave dressing field, as shown in Fig.~\ref{Fig2}(a).

Remarkably, we identified four primary loss spectrum structure far away $\Delta_{A}=0$, corresponding to the four main allowed transitions: $\ket{+}_{n}$ $\leftrightarrow$ $\ket{-}_{n+1}$, $\ket{+}_{n}$ $\leftrightarrow$ $\ket{0}_{n+1}$, $\ket{+}_{n+1}$ $\leftrightarrow$ $\ket{0}_{n}$, $\ket{+}_{n+1}$ $\leftrightarrow$ $\ket{-}_{n}$, as depicted in Fig.~\ref{Fig1}(c). Among these transitions, the $\ket{+}_{n}$ $\leftrightarrow$ $\ket{-}_{n+1}$ and $\ket{+}_{n+1}$ $\leftrightarrow$ $\ket{-}_{n}$ transitions, can be interpreted using the dressed state approach of two energy level (excited states $N=1$) with a Mollow triplet structure~\cite{CohenTannoudji1998}. The emitted fluorescence spectrum of Mollow triplet comprises a central transition at $\omega_{d}$ and two sidebands at $\omega_{d}\pm\Omega_{\text{eff}}$. Here, two loss peaks associated with two $\ket{+}_{n}$ $\leftrightarrow$ $\ket{-}_{n+1}$ and $\ket{+}_{n+1}$ $\leftrightarrow$ $\ket{-}_{n}$ transitions correspond to the two sidebands of the Mollow triplet. The effective coupling strength, denoted as $\Omega_{\text{eff}}=\sqrt{\Omega^{2}+\Delta^{2}}$, can be determined from the spectral distance between the two transition frequencies, $\Omega$ denotes the Rabi coupling strength of microwave dressing field on resonance $\Delta=0$. This provides a method to precisely measure the coupling strength $\Omega$ at high microwave power levels~\cite{zhang2024Phys.Rev.Lett.}. In the case of blue detuning dressing field, the transition of $\ket{+}_{n}$ $\leftrightarrow$ $\ket{-}_{n+1}$ has a greater dipole moment compared to $\ket{+}_{n+1}$ $\leftrightarrow$ $\ket{-}_{n}$ transition, resulting in a more substantial loss depth. Conversely, for the red detuning dressing field, the observation is reversed, with a more significant loss depth occurring at the transition of $\ket{+}_{n+1}$ $\leftrightarrow$ $\ket{-}_{n}$, as detailed in supplementary materials.

In addition, the hyperfine couplings in the first rotational state (the excited states $N\geq 2$) results in the formation of a dark state manifold $\ket{0}$ within the energy level system, which is a linear superpositions of multiple hyperfine states of the first rotational state, as illustrated in Fig.~\ref{Fig1}(c). We identified the presence of two prominent symmetric loss peaks at the two transition frequencies $\ket{+}_{n}$ $\leftrightarrow$ $\ket{0}_{n+1}$ and $\ket{+}_{n+1}$ $\leftrightarrow$ $\ket{0}_{n}$ in the dressed state spectroscopy. A more significant loss depth was observed at the transition $\ket{+}_{n}$ $\leftrightarrow$ $\ket{0}_{n+1}$ with $\Delta_{A}<0$, which is comparable to the transition $\ket{+}_{n}$ $\leftrightarrow$ $\ket{-}_{n+1}$, as depicted in Fig.~\ref{Fig2}(a). With the red detuned dressing field, the situation is reversed. There is a greater loss depth at the transition between $\ket{+}_{n+1}$ and $\ket{0}_{n}$ with $\Delta_{A}>0$, as described in the supplementary materials.

\begin{figure}[t]

\centerline{\includegraphics[width=3.4 in]{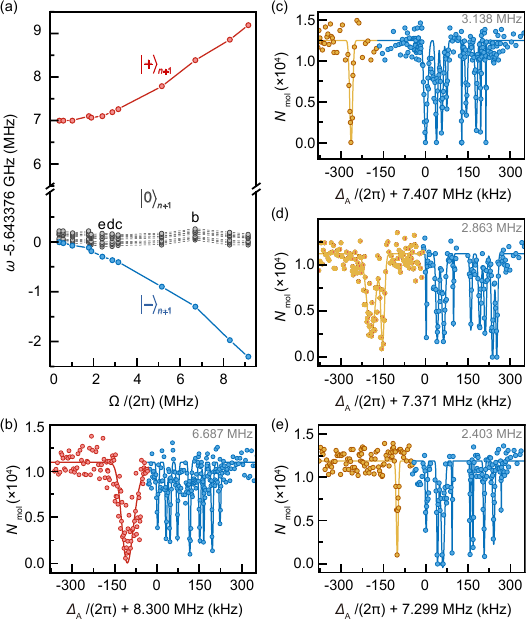}}
\caption{(Color online). \textbf{The energies of the two dressed states and hyperfine dark states obtained from dressed state spectroscopy and the resonance coupling between the dressed state and the tetratomic state.}
(a) Energies of the two dressed states and nine hyperfine dark states as a function of the coupling strength of the microwave dressing field, obtained from dressed-state spectroscopy.
(b) Typical dressed state spectroscopy at large microwave coupling strength of $2\pi\times6.687$ MHz, tetratomic molecules are associated solely via the association microwave field. The red line is a Gaussian fit to the data, corresponding to the tetratomic state.
(c, d, e) Typical spectroscopy at weak dressing field, where the association microwave field alone cannot associate tetratomic molecules (c, e), but association occurs through resonant coupling between the dressed state and the tetratomic state (d).
\label{Fig3} }
\end{figure}

\begin{figure}[t]
\centerline{\includegraphics[width=2.9 in]{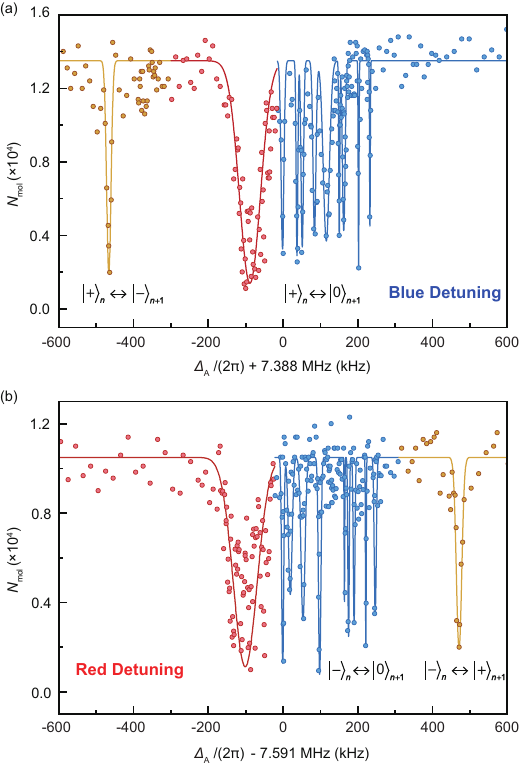}}
\caption{(Color online). \textbf{Association of tetratomic molecules under blue- and red-detuned dressing microwave fields}.
Loss spectra are compared for a dressing microwave field with a coupling strength of $\Omega = 2\pi \times 3.548\,\mathrm{MHz}$: (a) blue detuning of $\Delta = 2\pi \times 7\,\mathrm{MHz}$, and (b) red detuning of $\Delta = -2\pi \times 7\,\mathrm{MHz}$. The broad red loss feature corresponds to the association of tetratomic molecules with approximately equal binding energies under both detuning conditions. The narrow yellow peaks indicate transitions between the two dressed states $\ket{+}$ and $\ket{-}$. The narrow blue peaks represent the typical nine transitions between the dressed state and dark states manifolds.
\label{Fig4} }
\end{figure}

In the following, we focus on the detailed loss spectrum of the $\ket{+}_{n} \leftrightarrow \ket{0}_{n+1}$ transition, as shown in Fig.~\ref{Fig2}(b). Nine distinct loss peaks, highlighted by the blue lines, are clearly observed and can be attributed to transitions from the dressed state $\ket{+}_{n}$ to the dark states $\ket{0}_{n+1}$ of freely dressed molecules. Notably, an additional loss peak is observed near the first free-free dressed molecule state transitions $\ket{+}_{n}$ $\rightarrow$ $\ket{0}^{1}_{n+1}$, occurring at a frequency of about 100 kHz lower than the first transition line. This loss peak, characterized by a broader spectral width, can be distinctly differentiated from the loss associated with transitions to the dark states. The additional loss peak observed in the spectrum signals the formation of tetratomic molecules through microwave association, which arises from the combined contributions of several odd angular momentum tetramer states, as predicted theoretically in Section 3 of the Supplementary Information. These tetramer states have similar binding energies and short lifetimes, leading to broad loss features that render them experimentally indistinguishable. This process occurs when two free dipolar molecules, initially in the dressed state $\ket{+}$, transition into a microwave-dressing-induced tetramer state within the entrance channel $\ket{+0}$, which comprises a dark excited state $\ket{0}$ and a dressed state $\ket{+}$. This approach is analogous to the technique of associating weakly bound molecules near a Feshbach resonance through radio frequency driven transitions that alter atomic hyperfine states~\cite{PhysRevLett.97.120402,PhysRevLett.100.143201}. We extract the binding energy of the tetratomic molecules, approximately $h$ $\times$ 100 kHz, from the spectral distance between the loss peak of the tetratomic molecules and the initial free-molecule transition peak in the microwave-dressed spectroscopy.

The fermionic molecules are initially prepared in the same dressed state $\ket{+}$, such that the antisymmetrized two-body state of colliding molecules with momenta $\mathbf{p}$ and $\mathbf{q}$ is
$\frac{1}{\sqrt{2}}\bigl(\ket{\mathbf{p}}_{1}\ket{\mathbf{q}}_{2}-\ket{\mathbf{q}}_{1}\ket{\mathbf{p}}_{2}\bigr)\ket{+}_{1}\ket{+}_{2}$,
which represents $p$-wave interactions and contains only triplet spin components~\cite{Fu2013}.
Since the microwave field couples the initial state to the tetramer states without changing the momentum part, the corresponding final tetramer states in $\ket{+0}_{s}$ channel are
$\frac{1}{\sqrt{2}}\bigl(\ket{\mathbf{p}}_{1}\ket{\mathbf{q}}_{2}-\ket{\mathbf{q}}_{1}\ket{\mathbf{p}}_{2}\bigr)
\frac{1}{\sqrt{2}}\bigl(\ket{+}_{1}\ket{0}_{2}+\ket{0}_{1}\ket{+}_{2}\bigr)$,
where only the spin part is changed, retaining triplet character and ensuring a nonzero transition dipole moment from the initial state under microwave driving, as shown in Fig.~\ref{Fig1}(d). The properties of tetramer states are discussed in section 3 in supplementary information. Here, the angular momentum part $(l,m)$ of the tetramer state corresponds to odd angular momentum, as required by antisymmetrization of the momentum wave function. This case contrasts significantly with our previous work of coherently production of s-wave Feshbach molecules from a fully polarized Fermi gas by spin–orbit coupling of Raman process~\cite{Fu2013}.

We measure the dressed-state spectroscopy at a blue detuning of $\Delta = 2\pi \times 7~\mathrm{MHz}$ for various coupling strengths of the microwave dressing field. The energies of the two dressed states and the hyperfine dark states are extracted from the spectral spacings between the four observed loss peaks, as shown in Fig.~\ref{Fig3}(a). The dressing field plays a central role in the formation of bound tetratomic molecules. We observe that the tetratomic molecules state appears only when the dressing microwave coupling strength exceeds $\sim 2\pi\times3.546$ MHz, with a binding energy of approximately $h$ $\times$ 100 kHz that remains nearly constant over this range. At weaker dressing fields, tetratomic molecules cannot be associated. Remarkably, when the microwave field is tuned such that the dressed state becomes resonant with the tetratomic state, tetratomic molecules can be re-associated even under a very weak dressing field (see Figs.~\ref{Fig3}(c,d,e)). This enhancement arises because the near-resonant condition strongly amplifies the coupling between the free dressed state and the bound tetratomic state.


Finally, we also observe the association of tetratomic molecules under red-detuned microwave dressing field. In particular, we compare the loss spectra at a blue detuning of $\Delta=2\pi\times7$ MHz and a red detuning of $\Delta=2\pi\times -7$ MHz, utilizing the same coupling strength of $\Omega\sim2\pi\times3.548$ MHz, as shown in Fig.~\ref{Fig4}. Our measurements indicate that the loss features in both scenarios exhibit similar structures, and the binding energy of the tetratomic molecules formed under red detuning is approximately equivalent to that of those formed under blue detuning. We found that the microwave association of tetratomic molecules in our approach is not sensitive to the polarization of the microwave dressing field (see supplementary materials), which is significantly different with the electroassociation technique~\cite{chen2024Nature}.

In conclusion, we have demonstrated microwave spectroscopy of tetratomic molecules associated from microwave-dressed fermionic $^{23}$Na$^{40}$K, marking a first step toward the controlled production of ultracold tetratomic molecules. By applying a second microwave field, we precisely characterize the dressed diatomic molecules, resolve the complex hyperfine structure of the dark excited state $\ket{0}$, and determine the binding energy of the tetratomic molecules. The bound tetramer state can be employed to control interactions between diatomic molecules~\cite{PhysRevLett.121.163402} by tuning the strength and detuning of the association microwave field that couples the free scattering state of two dressed diatomic molecules to the bound tetramer state, in direct analogy to optical Feshbach resonances in neutral-atom systems~\cite{PhysRevLett.77.2913,PhysRevA.56.1486,PhysRevLett.85.4462}.

Microwave dressing not only provides precise control over the internal rotational states but also induces an effective shielding mechanism that suppresses inelastic collisions while preserving elastic dipolar interactions~\cite{schindewolf2022Nature,Bigagli2023Nat.Phys.,lin2023Phys.Rev.X,anderegg2021Science}. This microwave shielding plays a pivotal role in enhancing elastic collision rates and enabling efficient evaporative cooling—key ingredients for achieving quantum-degenerate gases of fermionic molecules~\cite{schindewolf2022Nature} and Bose–Einstein condensates of dipolar bosonic species~\cite{bigagli2024Naturea,shi2025}. The spectroscopic characterization and prospective controlled creation of tetramer states in such a microwave-dressed environment establish a promising platform for studying few-body physics and dipolar interactions in regimes where shielding enhances both collisional stability and state selectivity.

Our method provides a general framework that can be extended to other ultracold diatomic molecular systems for the creation and control of weakly bound polyatomic states. The realization of weakly bound tetratomic molecules from microwave-dressed polar molecules offers a clean and tunable platform to investigate quantum few-body phenomena beyond the three-body regime, including four-body resonances, bound-state interference, and the emergence of correlated few-body dynamics. Furthermore, the ability to resolve and manipulate hyperfine-resolved tetramer states opens new prospects for exploring internal-state-controlled quantum chemistry and energy transfer processes in ultracold polyatomic systems.

\begin{acknowledgments}

This research is supported by National Key Research and Development Program of China (Grants Nos.~2022YFA1404101, and~2021YFA1401700), Innovation Program for Quantum Science and Technology (Grants No.~2021ZD0302003), National Natural Science Foundation of China (Grants Nos.~12034011, ~U23A6004, ~12488301, ~12474266, ~12474252, ~12374245, ~12322409).

\end{acknowledgments}

\bibliography{references}

\begin{thebibliography}{60}%
\makeatletter
\providecommand \@ifxundefined [1]{%
 \@ifx{#1\undefined}
}%
\providecommand \@ifnum [1]{%
 \ifnum #1\expandafter \@firstoftwo
 \else \expandafter \@secondoftwo
 \fi
}%
\providecommand \@ifx [1]{%
 \ifx #1\expandafter \@firstoftwo
 \else \expandafter \@secondoftwo
 \fi
}%
\providecommand \natexlab [1]{#1}%
\providecommand \enquote  [1]{``#1''}%
\providecommand \bibnamefont  [1]{#1}%
\providecommand \bibfnamefont [1]{#1}%
\providecommand \citenamefont [1]{#1}%
\providecommand \href@noop [0]{\@secondoftwo}%
\providecommand \href [0]{\begingroup \@sanitize@url \@href}%
\providecommand \@href[1]{\@@startlink{#1}\@@href}%
\providecommand \@@href[1]{\endgroup#1\@@endlink}%
\providecommand \@sanitize@url [0]{\catcode `\\12\catcode `\$12\catcode
  `\&12\catcode `\#12\catcode `\^12\catcode `\_12\catcode `\%12\relax}%
\providecommand \@@startlink[1]{}%
\providecommand \@@endlink[0]{}%
\providecommand \url  [0]{\begingroup\@sanitize@url \@url }%
\providecommand \@url [1]{\endgroup\@href {#1}{\urlprefix }}%
\providecommand \urlprefix  [0]{URL }%
\providecommand \Eprint [0]{\href }%
\providecommand \doibase [0]{https://doi.org/}%
\providecommand \selectlanguage [0]{\@gobble}%
\providecommand \bibinfo  [0]{\@secondoftwo}%
\providecommand \bibfield  [0]{\@secondoftwo}%
\providecommand \translation [1]{[#1]}%
\providecommand \BibitemOpen [0]{}%
\providecommand \bibitemStop [0]{}%
\providecommand \bibitemNoStop [0]{.\EOS\space}%
\providecommand \EOS [0]{\spacefactor3000\relax}%
\providecommand \BibitemShut  [1]{\csname bibitem#1\endcsname}%
\let\auto@bib@innerbib\@empty
\bibitem [{\citenamefont {Krems}(2008)}]{Krems2008}%
  \BibitemOpen
  \bibfield  {author} {\bibinfo {author} {\bibfnamefont {R.~V.}\ \bibnamefont
  {Krems}},\ }\bibfield  {title} {\bibinfo {title} {Cold controlled
  chemistry},\ }\href {https://doi.org/10.1039/b802322k} {\bibfield  {journal}
  {\bibinfo  {journal} {Phys. Chem. Chem. Phys.}\ }\textbf {\bibinfo {volume}
  {10}},\ \bibinfo {pages} {4079} (\bibinfo {year} {2008})}\BibitemShut
  {NoStop}%
\bibitem [{\citenamefont {Liu}\ and\ \citenamefont
  {Ni}(2022)}]{liu2022Annu.Rev.Phys.Chem.}%
  \BibitemOpen
  \bibfield  {author} {\bibinfo {author} {\bibfnamefont {Y.}~\bibnamefont
  {Liu}}\ and\ \bibinfo {author} {\bibfnamefont {K.-K.}\ \bibnamefont {Ni}},\
  }\bibfield  {title} {\bibinfo {title} {Bimolecular {{Chemistry}} in the
  {{Ultracold Regime}}},\ }\href
  {https://doi.org/10.1146/annurev-physchem-090419-043244} {\bibfield
  {journal} {\bibinfo  {journal} {Annu. Rev. Phys. Chem.}\ }\textbf {\bibinfo
  {volume} {73}},\ \bibinfo {pages} {73} (\bibinfo {year} {2022})}\BibitemShut
  {NoStop}%
\bibitem [{\citenamefont {Karman}\ \emph {et~al.}(2024)\citenamefont {Karman},
  \citenamefont {Tomza},\ and\ \citenamefont
  {{P{\'e}rez-R{\'i}os}}}]{karman2024Nat.Phys.a}%
  \BibitemOpen
  \bibfield  {author} {\bibinfo {author} {\bibfnamefont {T.}~\bibnamefont
  {Karman}}, \bibinfo {author} {\bibfnamefont {M.}~\bibnamefont {Tomza}},\ and\
  \bibinfo {author} {\bibfnamefont {J.}~\bibnamefont {{P{\'e}rez-R{\'i}os}}},\
  }\bibfield  {title} {\bibinfo {title} {Ultracold chemistry as a testbed for
  few-body physics},\ }\href {https://doi.org/10.1038/s41567-024-02467-3}
  {\bibfield  {journal} {\bibinfo  {journal} {Nat. Phys.}\ }\textbf {\bibinfo
  {volume} {20}},\ \bibinfo {pages} {722} (\bibinfo {year} {2024})}\BibitemShut
  {NoStop}%
\bibitem [{\citenamefont {DeMille}\ \emph {et~al.}(2008)\citenamefont
  {DeMille}, \citenamefont {Sainis}, \citenamefont {Sage}, \citenamefont
  {Bergeman}, \citenamefont {Kotochigova},\ and\ \citenamefont
  {Tiesinga}}]{PhysRevLett.100.043202}%
  \BibitemOpen
  \bibfield  {author} {\bibinfo {author} {\bibfnamefont {D.}~\bibnamefont
  {DeMille}}, \bibinfo {author} {\bibfnamefont {S.}~\bibnamefont {Sainis}},
  \bibinfo {author} {\bibfnamefont {J.}~\bibnamefont {Sage}}, \bibinfo {author}
  {\bibfnamefont {T.}~\bibnamefont {Bergeman}}, \bibinfo {author}
  {\bibfnamefont {S.}~\bibnamefont {Kotochigova}},\ and\ \bibinfo {author}
  {\bibfnamefont {E.}~\bibnamefont {Tiesinga}},\ }\bibfield  {title} {\bibinfo
  {title} {Enhanced sensitivity to variation of ${m}_{e}/{m}_{p}$ in molecular
  spectra},\ }\href {https://doi.org/10.1103/PhysRevLett.100.043202} {\bibfield
   {journal} {\bibinfo  {journal} {Phys. Rev. Lett.}\ }\textbf {\bibinfo
  {volume} {100}},\ \bibinfo {pages} {043202} (\bibinfo {year}
  {2008})}\BibitemShut {NoStop}%
\bibitem [{\citenamefont {Collaboration}\ \emph {et~al.}(2014)\citenamefont
  {Collaboration}, \citenamefont {Baron}, \citenamefont {Campbell},
  \citenamefont {DeMille}, \citenamefont {Doyle}, \citenamefont {Gabrielse},
  \citenamefont {Gurevich}, \citenamefont {Hess}, \citenamefont {Hutzler},
  \citenamefont {Kirilov}, \citenamefont {Kozyryev}, \citenamefont {O’Leary},
  \citenamefont {Panda}, \citenamefont {Parsons}, \citenamefont {Petrik},
  \citenamefont {Spaun}, \citenamefont {Vutha},\ and\ \citenamefont
  {West}}]{ACME2014}%
  \BibitemOpen
  \bibfield  {author} {\bibinfo {author} {\bibfnamefont {T.~A.}\ \bibnamefont
  {Collaboration}}, \bibinfo {author} {\bibfnamefont {J.}~\bibnamefont
  {Baron}}, \bibinfo {author} {\bibfnamefont {W.~C.}\ \bibnamefont {Campbell}},
  \bibinfo {author} {\bibfnamefont {D.}~\bibnamefont {DeMille}}, \bibinfo
  {author} {\bibfnamefont {J.~M.}\ \bibnamefont {Doyle}}, \bibinfo {author}
  {\bibfnamefont {G.}~\bibnamefont {Gabrielse}}, \bibinfo {author}
  {\bibfnamefont {Y.~V.}\ \bibnamefont {Gurevich}}, \bibinfo {author}
  {\bibfnamefont {P.~W.}\ \bibnamefont {Hess}}, \bibinfo {author}
  {\bibfnamefont {N.~R.}\ \bibnamefont {Hutzler}}, \bibinfo {author}
  {\bibfnamefont {E.}~\bibnamefont {Kirilov}}, \bibinfo {author} {\bibfnamefont
  {I.}~\bibnamefont {Kozyryev}}, \bibinfo {author} {\bibfnamefont {B.~R.}\
  \bibnamefont {O’Leary}}, \bibinfo {author} {\bibfnamefont {C.~D.}\
  \bibnamefont {Panda}}, \bibinfo {author} {\bibfnamefont {M.~F.}\ \bibnamefont
  {Parsons}}, \bibinfo {author} {\bibfnamefont {E.~S.}\ \bibnamefont {Petrik}},
  \bibinfo {author} {\bibfnamefont {B.}~\bibnamefont {Spaun}}, \bibinfo
  {author} {\bibfnamefont {A.~C.}\ \bibnamefont {Vutha}},\ and\ \bibinfo
  {author} {\bibfnamefont {A.~D.}\ \bibnamefont {West}},\ }\bibfield  {title}
  {\bibinfo {title} {Order of magnitude smaller limit on the electric dipole
  moment of the electron},\ }\href {https://doi.org/10.1126/science.1248213}
  {\bibfield  {journal} {\bibinfo  {journal} {Science}\ }\textbf {\bibinfo
  {volume} {343}},\ \bibinfo {pages} {269} (\bibinfo {year}
  {2014})}\BibitemShut {NoStop}%
\bibitem [{\citenamefont {Ye}\ and\ \citenamefont
  {Zoller}(2024)}]{ye2024Phys.Rev.Lett.}%
  \BibitemOpen
  \bibfield  {author} {\bibinfo {author} {\bibfnamefont {J.}~\bibnamefont
  {Ye}}\ and\ \bibinfo {author} {\bibfnamefont {P.}~\bibnamefont {Zoller}},\
  }\bibfield  {title} {\bibinfo {title} {Essay: {{Quantum Sensing}} with
  {{Atomic}}, {{Molecular}}, and {{Optical Platforms}} for {{Fundamental
  Physics}}},\ }\href {https://doi.org/10.1103/PhysRevLett.132.190001}
  {\bibfield  {journal} {\bibinfo  {journal} {Phys. Rev. Lett.}\ }\textbf
  {\bibinfo {volume} {132}},\ \bibinfo {pages} {190001} (\bibinfo {year}
  {2024})}\BibitemShut {NoStop}%
\bibitem [{\citenamefont {DeMille}\ \emph {et~al.}(2024)\citenamefont
  {DeMille}, \citenamefont {Hutzler}, \citenamefont {Rey},\ and\ \citenamefont
  {Zelevinsky}}]{demille2024Nat.Phys.}%
  \BibitemOpen
  \bibfield  {author} {\bibinfo {author} {\bibfnamefont {D.}~\bibnamefont
  {DeMille}}, \bibinfo {author} {\bibfnamefont {N.~R.}\ \bibnamefont
  {Hutzler}}, \bibinfo {author} {\bibfnamefont {A.~M.}\ \bibnamefont {Rey}},\
  and\ \bibinfo {author} {\bibfnamefont {T.}~\bibnamefont {Zelevinsky}},\
  }\bibfield  {title} {\bibinfo {title} {Quantum sensing and metrology for
  fundamental physics with molecules},\ }\href
  {https://doi.org/10.1038/s41567-024-02499-9} {\bibfield  {journal} {\bibinfo
  {journal} {Nat. Phys.}\ }\textbf {\bibinfo {volume} {20}},\ \bibinfo {pages}
  {741} (\bibinfo {year} {2024})}\BibitemShut {NoStop}%
\bibitem [{\citenamefont {Moses}\ \emph {et~al.}(2017)\citenamefont {Moses},
  \citenamefont {Covey}, \citenamefont {Miecnikowski}, \citenamefont {Jin},\
  and\ \citenamefont {Ye}}]{moses2017NaturePhys}%
  \BibitemOpen
  \bibfield  {author} {\bibinfo {author} {\bibfnamefont {S.~A.}\ \bibnamefont
  {Moses}}, \bibinfo {author} {\bibfnamefont {J.~P.}\ \bibnamefont {Covey}},
  \bibinfo {author} {\bibfnamefont {M.~T.}\ \bibnamefont {Miecnikowski}},
  \bibinfo {author} {\bibfnamefont {D.~S.}\ \bibnamefont {Jin}},\ and\ \bibinfo
  {author} {\bibfnamefont {J.}~\bibnamefont {Ye}},\ }\bibfield  {title}
  {\bibinfo {title} {New frontiers for quantum gases of polar molecules},\
  }\href {https://doi.org/10.1038/nphys3985} {\bibfield  {journal} {\bibinfo
  {journal} {Nat. Phys.}\ }\textbf {\bibinfo {volume} {13}},\ \bibinfo {pages}
  {13} (\bibinfo {year} {2017})}\BibitemShut {NoStop}%
\bibitem [{\citenamefont {Kruckenhauser}\ \emph {et~al.}(2020)\citenamefont
  {Kruckenhauser}, \citenamefont {Sieberer}, \citenamefont {De~Marco},
  \citenamefont {Li}, \citenamefont {Matsuda}, \citenamefont {Tobias},
  \citenamefont {Valtolina}, \citenamefont {Ye}, \citenamefont {Rey},
  \citenamefont {Baranov},\ and\ \citenamefont {Zoller}}]{PhysRevA.102.023320}%
  \BibitemOpen
  \bibfield  {author} {\bibinfo {author} {\bibfnamefont {A.}~\bibnamefont
  {Kruckenhauser}}, \bibinfo {author} {\bibfnamefont {L.~M.}\ \bibnamefont
  {Sieberer}}, \bibinfo {author} {\bibfnamefont {L.}~\bibnamefont {De~Marco}},
  \bibinfo {author} {\bibfnamefont {J.-R.}\ \bibnamefont {Li}}, \bibinfo
  {author} {\bibfnamefont {K.}~\bibnamefont {Matsuda}}, \bibinfo {author}
  {\bibfnamefont {W.~G.}\ \bibnamefont {Tobias}}, \bibinfo {author}
  {\bibfnamefont {G.}~\bibnamefont {Valtolina}}, \bibinfo {author}
  {\bibfnamefont {J.}~\bibnamefont {Ye}}, \bibinfo {author} {\bibfnamefont
  {A.~M.}\ \bibnamefont {Rey}}, \bibinfo {author} {\bibfnamefont {M.~A.}\
  \bibnamefont {Baranov}},\ and\ \bibinfo {author} {\bibfnamefont
  {P.}~\bibnamefont {Zoller}},\ }\bibfield  {title} {\bibinfo {title} {Quantum
  many-body physics with ultracold polar molecules: Nanostructured potential
  barriers and interactions},\ }\href
  {https://doi.org/10.1103/PhysRevA.102.023320} {\bibfield  {journal} {\bibinfo
   {journal} {Phys. Rev. A}\ }\textbf {\bibinfo {volume} {102}},\ \bibinfo
  {pages} {023320} (\bibinfo {year} {2020})}\BibitemShut {NoStop}%
\bibitem [{\citenamefont {Cornish}\ \emph {et~al.}(2024)\citenamefont
  {Cornish}, \citenamefont {Tarbutt},\ and\ \citenamefont
  {Hazzard}}]{Cornish2024}%
  \BibitemOpen
  \bibfield  {author} {\bibinfo {author} {\bibfnamefont {S.~L.}\ \bibnamefont
  {Cornish}}, \bibinfo {author} {\bibfnamefont {M.~R.}\ \bibnamefont
  {Tarbutt}},\ and\ \bibinfo {author} {\bibfnamefont {K.~R.~A.}\ \bibnamefont
  {Hazzard}},\ }\bibfield  {title} {\bibinfo {title} {Quantum computation and
  quantum simulation with ultracold molecules},\ }\href
  {https://doi.org/10.1038/s41567-024-02453-9} {\bibfield  {journal} {\bibinfo
  {journal} {Nat. Phys.}\ }\textbf {\bibinfo {volume} {20}},\ \bibinfo {pages}
  {730–740} (\bibinfo {year} {2024})}\BibitemShut {NoStop}%
\bibitem [{\citenamefont {Balakrishnan}(2016)}]{Balakrishnan2016}%
  \BibitemOpen
  \bibfield  {author} {\bibinfo {author} {\bibfnamefont {N.}~\bibnamefont
  {Balakrishnan}},\ }\bibfield  {title} {\bibinfo {title} {Perspective:
  Ultracold molecules and the dawn of cold controlled chemistry},\ }\href
  {https://doi.org/10.1063/1.4964096} {\bibfield  {journal} {\bibinfo
  {journal} {J. Chem. Phys.}\ }\textbf {\bibinfo {volume} {145}},\ \bibinfo
  {pages} {150901} (\bibinfo {year} {2016})}\BibitemShut {NoStop}%
\bibitem [{\citenamefont {Safronova}\ \emph {et~al.}(2018)\citenamefont
  {Safronova}, \citenamefont {Budker}, \citenamefont {DeMille}, \citenamefont
  {Kimball}, \citenamefont {Derevianko},\ and\ \citenamefont
  {Clark}}]{safronova2018Rev.Mod.Phys.}%
  \BibitemOpen
  \bibfield  {author} {\bibinfo {author} {\bibfnamefont {M.~S.}\ \bibnamefont
  {Safronova}}, \bibinfo {author} {\bibfnamefont {D.}~\bibnamefont {Budker}},
  \bibinfo {author} {\bibfnamefont {D.}~\bibnamefont {DeMille}}, \bibinfo
  {author} {\bibfnamefont {D.~F.~J.}\ \bibnamefont {Kimball}}, \bibinfo
  {author} {\bibfnamefont {A.}~\bibnamefont {Derevianko}},\ and\ \bibinfo
  {author} {\bibfnamefont {C.~W.}\ \bibnamefont {Clark}},\ }\bibfield  {title}
  {\bibinfo {title} {Search for new physics with atoms and molecules},\ }\href
  {https://doi.org/10.1103/RevModPhys.90.025008} {\bibfield  {journal}
  {\bibinfo  {journal} {Rev. Mod. Phys.}\ }\textbf {\bibinfo {volume} {90}},\
  \bibinfo {pages} {025008} (\bibinfo {year} {2018})}\BibitemShut {NoStop}%
\bibitem [{\citenamefont {Gacesa}\ \emph {et~al.}(2021)\citenamefont {Gacesa},
  \citenamefont {Byrd}, \citenamefont {Smucker}, \citenamefont {Montgomery},\
  and\ \citenamefont {C{\^o}t{\'e}}}]{gacesa2021Phys.Rev.Research}%
  \BibitemOpen
  \bibfield  {author} {\bibinfo {author} {\bibfnamefont {M.}~\bibnamefont
  {Gacesa}}, \bibinfo {author} {\bibfnamefont {J.~N.}\ \bibnamefont {Byrd}},
  \bibinfo {author} {\bibfnamefont {J.}~\bibnamefont {Smucker}}, \bibinfo
  {author} {\bibfnamefont {J.~A.}\ \bibnamefont {Montgomery}},\ and\ \bibinfo
  {author} {\bibfnamefont {R.}~\bibnamefont {C{\^o}t{\'e}}},\ }\bibfield
  {title} {\bibinfo {title} {Photoassociation of ultracold long-range
  polyatomic molecules},\ }\href
  {https://doi.org/10.1103/PhysRevResearch.3.023163} {\bibfield  {journal}
  {\bibinfo  {journal} {Phys. Rev. Research}\ }\textbf {\bibinfo {volume}
  {3}},\ \bibinfo {pages} {023163} (\bibinfo {year} {2021})}\BibitemShut
  {NoStop}%
\bibitem [{\citenamefont {Doyle}\ \emph {et~al.}(2022)\citenamefont {Doyle},
  \citenamefont {Augenbraun},\ and\ \citenamefont {Lasner}}]{Doyle2022}%
  \BibitemOpen
  \bibfield  {author} {\bibinfo {author} {\bibfnamefont {J.~M.}\ \bibnamefont
  {Doyle}}, \bibinfo {author} {\bibfnamefont {B.~L.}\ \bibnamefont
  {Augenbraun}},\ and\ \bibinfo {author} {\bibfnamefont {Z.~D.}\ \bibnamefont
  {Lasner}},\ }\bibfield  {title} {\bibinfo {title} {Ultracold polyatomic
  molecules for quantum science and precision measurements},\ }in\ \href
  {https://doi.org/10.7566/jpscp.37.011004} {\emph {\bibinfo {booktitle}
  {Proceedings of the 24th International Spin Symposium (SPIN2021)}}}\
  (\bibinfo  {publisher} {Journal of the Physical Society of Japan},\ \bibinfo
  {year} {2022})\BibitemShut {NoStop}%
\bibitem [{\citenamefont {Yang}\ \emph {et~al.}(2022)\citenamefont {Yang},
  \citenamefont {Cao}, \citenamefont {Su}, \citenamefont {Rui}, \citenamefont
  {Zhao},\ and\ \citenamefont {Pan}}]{yang2022Science}%
  \BibitemOpen
  \bibfield  {author} {\bibinfo {author} {\bibfnamefont {H.}~\bibnamefont
  {Yang}}, \bibinfo {author} {\bibfnamefont {J.}~\bibnamefont {Cao}}, \bibinfo
  {author} {\bibfnamefont {Z.}~\bibnamefont {Su}}, \bibinfo {author}
  {\bibfnamefont {J.}~\bibnamefont {Rui}}, \bibinfo {author} {\bibfnamefont
  {B.}~\bibnamefont {Zhao}},\ and\ \bibinfo {author} {\bibfnamefont {J.-W.}\
  \bibnamefont {Pan}},\ }\bibfield  {title} {\bibinfo {title} {Creation of an
  ultracold gas of triatomic molecules from an atom--diatomic molecule
  mixture},\ }\href {https://doi.org/10.1126/science.ade6307} {\bibfield
  {journal} {\bibinfo  {journal} {Science}\ }\textbf {\bibinfo {volume}
  {378}},\ \bibinfo {pages} {1009} (\bibinfo {year} {2022})}\BibitemShut
  {NoStop}%
\bibitem [{\citenamefont {Chen}\ \emph {et~al.}(2024)\citenamefont {Chen},
  \citenamefont {Biswas}, \citenamefont {Eppelt}, \citenamefont {Schindewolf},
  \citenamefont {Deng}, \citenamefont {Shi}, \citenamefont {Yi}, \citenamefont
  {Hilker}, \citenamefont {Bloch},\ and\ \citenamefont {Luo}}]{chen2024Nature}%
  \BibitemOpen
  \bibfield  {author} {\bibinfo {author} {\bibfnamefont {X.-Y.}\ \bibnamefont
  {Chen}}, \bibinfo {author} {\bibfnamefont {S.}~\bibnamefont {Biswas}},
  \bibinfo {author} {\bibfnamefont {S.}~\bibnamefont {Eppelt}}, \bibinfo
  {author} {\bibfnamefont {A.}~\bibnamefont {Schindewolf}}, \bibinfo {author}
  {\bibfnamefont {F.}~\bibnamefont {Deng}}, \bibinfo {author} {\bibfnamefont
  {T.}~\bibnamefont {Shi}}, \bibinfo {author} {\bibfnamefont {S.}~\bibnamefont
  {Yi}}, \bibinfo {author} {\bibfnamefont {T.~A.}\ \bibnamefont {Hilker}},
  \bibinfo {author} {\bibfnamefont {I.}~\bibnamefont {Bloch}},\ and\ \bibinfo
  {author} {\bibfnamefont {X.-Y.}\ \bibnamefont {Luo}},\ }\bibfield  {title}
  {\bibinfo {title} {Ultracold field-linked tetratomic molecules},\ }\href
  {https://doi.org/10.1038/s41586-023-06986-6} {\bibfield  {journal} {\bibinfo
  {journal} {Nature}\ }\textbf {\bibinfo {volume} {626}},\ \bibinfo {pages}
  {283–287} (\bibinfo {year} {2024})}\BibitemShut {NoStop}%
\bibitem [{\citenamefont {B\"uchler}\ \emph {et~al.}(2007)\citenamefont
  {B\"uchler}, \citenamefont {Demler}, \citenamefont {Lukin}, \citenamefont
  {Micheli}, \citenamefont {Prokof'ev}, \citenamefont {Pupillo},\ and\
  \citenamefont {Zoller}}]{uchlerPhysRevLett.98.2007}%
  \BibitemOpen
  \bibfield  {author} {\bibinfo {author} {\bibfnamefont {H.~P.}\ \bibnamefont
  {B\"uchler}}, \bibinfo {author} {\bibfnamefont {E.}~\bibnamefont {Demler}},
  \bibinfo {author} {\bibfnamefont {M.}~\bibnamefont {Lukin}}, \bibinfo
  {author} {\bibfnamefont {A.}~\bibnamefont {Micheli}}, \bibinfo {author}
  {\bibfnamefont {N.}~\bibnamefont {Prokof'ev}}, \bibinfo {author}
  {\bibfnamefont {G.}~\bibnamefont {Pupillo}},\ and\ \bibinfo {author}
  {\bibfnamefont {P.}~\bibnamefont {Zoller}},\ }\bibfield  {title} {\bibinfo
  {title} {Strongly correlated 2d quantum phases with cold polar molecules:
  Controlling the shape of the interaction potential},\ }\href
  {https://doi.org/10.1103/PhysRevLett.98.060404} {\bibfield  {journal}
  {\bibinfo  {journal} {Phys. Rev. Lett.}\ }\textbf {\bibinfo {volume} {98}},\
  \bibinfo {pages} {060404} (\bibinfo {year} {2007})}\BibitemShut {NoStop}%
\bibitem [{\citenamefont {Cooper}\ and\ \citenamefont
  {Shlyapnikov}(2009)}]{CooperPhysRevLett.103.2009}%
  \BibitemOpen
  \bibfield  {author} {\bibinfo {author} {\bibfnamefont {N.~R.}\ \bibnamefont
  {Cooper}}\ and\ \bibinfo {author} {\bibfnamefont {G.~V.}\ \bibnamefont
  {Shlyapnikov}},\ }\bibfield  {title} {\bibinfo {title} {Stable topological
  superfluid phase of ultracold polar fermionic molecules},\ }\href
  {https://doi.org/10.1103/PhysRevLett.103.155302} {\bibfield  {journal}
  {\bibinfo  {journal} {Phys. Rev. Lett.}\ }\textbf {\bibinfo {volume} {103}},\
  \bibinfo {pages} {155302} (\bibinfo {year} {2009})}\BibitemShut {NoStop}%
\bibitem [{\citenamefont {Levinsen}\ \emph {et~al.}(2011)\citenamefont
  {Levinsen}, \citenamefont {Cooper},\ and\ \citenamefont
  {Shlyapnikov}}]{LevinsenPhysRevA.84.2011}%
  \BibitemOpen
  \bibfield  {author} {\bibinfo {author} {\bibfnamefont {J.}~\bibnamefont
  {Levinsen}}, \bibinfo {author} {\bibfnamefont {N.~R.}\ \bibnamefont
  {Cooper}},\ and\ \bibinfo {author} {\bibfnamefont {G.~V.}\ \bibnamefont
  {Shlyapnikov}},\ }\bibfield  {title} {\bibinfo {title} {Topological
  ${p}_{x}+{\mathit{ip}}_{y}$ superfluid phase of fermionic polar molecules},\
  }\href {https://doi.org/10.1103/PhysRevA.84.013603} {\bibfield  {journal}
  {\bibinfo  {journal} {Phys. Rev. A}\ }\textbf {\bibinfo {volume} {84}},\
  \bibinfo {pages} {013603} (\bibinfo {year} {2011})}\BibitemShut {NoStop}%
\bibitem [{\citenamefont {Yan}\ \emph {et~al.}(2020)\citenamefont {Yan},
  \citenamefont {Park}, \citenamefont {Ni}, \citenamefont {Loh}, \citenamefont
  {Will}, \citenamefont {Karman},\ and\ \citenamefont
  {Zwierlein}}]{yanPhysRevLett.125.2020}%
  \BibitemOpen
  \bibfield  {author} {\bibinfo {author} {\bibfnamefont {Z.~Z.}\ \bibnamefont
  {Yan}}, \bibinfo {author} {\bibfnamefont {J.~W.}\ \bibnamefont {Park}},
  \bibinfo {author} {\bibfnamefont {Y.}~\bibnamefont {Ni}}, \bibinfo {author}
  {\bibfnamefont {H.}~\bibnamefont {Loh}}, \bibinfo {author} {\bibfnamefont
  {S.}~\bibnamefont {Will}}, \bibinfo {author} {\bibfnamefont {T.}~\bibnamefont
  {Karman}},\ and\ \bibinfo {author} {\bibfnamefont {M.}~\bibnamefont
  {Zwierlein}},\ }\bibfield  {title} {\bibinfo {title} {Resonant dipolar
  collisions of ultracold molecules induced by microwave dressing},\ }\href
  {https://doi.org/10.1103/PhysRevLett.125.063401} {\bibfield  {journal}
  {\bibinfo  {journal} {Phys. Rev. Lett.}\ }\textbf {\bibinfo {volume} {125}},\
  \bibinfo {pages} {063401} (\bibinfo {year} {2020})}\BibitemShut {NoStop}%
\bibitem [{\citenamefont {Hutzler}(2020)}]{hutzler2020QuantumSci.Technol.}%
  \BibitemOpen
  \bibfield  {author} {\bibinfo {author} {\bibfnamefont {N.~R.}\ \bibnamefont
  {Hutzler}},\ }\bibfield  {title} {\bibinfo {title} {Polyatomic molecules as
  quantum sensors for fundamental physics},\ }\href
  {https://doi.org/10.1088/2058-9565/abb9c5} {\bibfield  {journal} {\bibinfo
  {journal} {Quantum Sci. Technol.}\ }\textbf {\bibinfo {volume} {5}},\
  \bibinfo {pages} {044011} (\bibinfo {year} {2020})}\BibitemShut {NoStop}%
\bibitem [{\citenamefont {Vilas}\ \emph {et~al.}(2024)\citenamefont {Vilas},
  \citenamefont {Robichaud}, \citenamefont {Hallas}, \citenamefont {Li},
  \citenamefont {Anderegg},\ and\ \citenamefont {Doyle}}]{vilas2024Nature}%
  \BibitemOpen
  \bibfield  {author} {\bibinfo {author} {\bibfnamefont {N.~B.}\ \bibnamefont
  {Vilas}}, \bibinfo {author} {\bibfnamefont {P.}~\bibnamefont {Robichaud}},
  \bibinfo {author} {\bibfnamefont {C.}~\bibnamefont {Hallas}}, \bibinfo
  {author} {\bibfnamefont {G.~K.}\ \bibnamefont {Li}}, \bibinfo {author}
  {\bibfnamefont {L.}~\bibnamefont {Anderegg}},\ and\ \bibinfo {author}
  {\bibfnamefont {J.~M.}\ \bibnamefont {Doyle}},\ }\bibfield  {title} {\bibinfo
  {title} {An optical tweezer array of ultracold polyatomic molecules},\ }\href
  {https://doi.org/10.1038/s41586-024-07199-1} {\bibfield  {journal} {\bibinfo
  {journal} {Nature}\ }\textbf {\bibinfo {volume} {628}},\ \bibinfo {pages}
  {282} (\bibinfo {year} {2024})}\BibitemShut {NoStop}%
\bibitem [{\citenamefont {Christianen}\ \emph {et~al.}(2019)\citenamefont
  {Christianen}, \citenamefont {Karman}, \citenamefont
  {{Vargas-Hern{\'a}ndez}}, \citenamefont {Groenenboom},\ and\ \citenamefont
  {Krems}}]{christianen2019J.Chem.Phys.}%
  \BibitemOpen
  \bibfield  {author} {\bibinfo {author} {\bibfnamefont {A.}~\bibnamefont
  {Christianen}}, \bibinfo {author} {\bibfnamefont {T.}~\bibnamefont {Karman}},
  \bibinfo {author} {\bibfnamefont {R.~A.}\ \bibnamefont
  {{Vargas-Hern{\'a}ndez}}}, \bibinfo {author} {\bibfnamefont {G.~C.}\
  \bibnamefont {Groenenboom}},\ and\ \bibinfo {author} {\bibfnamefont {R.~V.}\
  \bibnamefont {Krems}},\ }\bibfield  {title} {\bibinfo {title}
  {Six-dimensional potential energy surface for {{NaK}}--{{NaK}} collisions:
  {{Gaussian}} process representation with correct asymptotic form},\ }\href
  {https://doi.org/10.1063/1.5082740} {\bibfield  {journal} {\bibinfo
  {journal} {J. Chem. Phys.}\ }\textbf {\bibinfo {volume} {150}},\ \bibinfo
  {pages} {064106} (\bibinfo {year} {2019})}\BibitemShut {NoStop}%
\bibitem [{\citenamefont {Unke}\ \emph {et~al.}(2020)\citenamefont {Unke},
  \citenamefont {Koner}, \citenamefont {Patra}, \citenamefont {K\"{a}ser},\
  and\ \citenamefont {Meuwly}}]{Unke2020}%
  \BibitemOpen
  \bibfield  {author} {\bibinfo {author} {\bibfnamefont {O.~T.}\ \bibnamefont
  {Unke}}, \bibinfo {author} {\bibfnamefont {D.}~\bibnamefont {Koner}},
  \bibinfo {author} {\bibfnamefont {S.}~\bibnamefont {Patra}}, \bibinfo
  {author} {\bibfnamefont {S.}~\bibnamefont {K\"{a}ser}},\ and\ \bibinfo
  {author} {\bibfnamefont {M.}~\bibnamefont {Meuwly}},\ }\bibfield  {title}
  {\bibinfo {title} {High-dimensional potential energy surfaces for molecular
  simulations: from empiricism to machine learning},\ }\href
  {https://doi.org/10.1088/2632-2153/ab5922} {\bibfield  {journal} {\bibinfo
  {journal} {Machine Learning: Science and Technology}\ }\textbf {\bibinfo
  {volume} {1}},\ \bibinfo {pages} {013001} (\bibinfo {year}
  {2020})}\BibitemShut {NoStop}%
\bibitem [{\citenamefont {Wheatley}\ \emph {et~al.}(2023)\citenamefont
  {Wheatley}, \citenamefont {Garberoglio},\ and\ \citenamefont
  {Harvey}}]{Wheatley2023}%
  \BibitemOpen
  \bibfield  {author} {\bibinfo {author} {\bibfnamefont {R.~J.}\ \bibnamefont
  {Wheatley}}, \bibinfo {author} {\bibfnamefont {G.}~\bibnamefont
  {Garberoglio}},\ and\ \bibinfo {author} {\bibfnamefont {A.~H.}\ \bibnamefont
  {Harvey}},\ }\bibfield  {title} {\bibinfo {title} {Four-body nonadditive
  potential energy surface and the fourth virial coefficient of helium},\
  }\href {https://doi.org/10.1021/acs.jced.3c00578} {\bibfield  {journal}
  {\bibinfo  {journal} {J. Chem. Eng. Data}\ }\textbf {\bibinfo {volume}
  {68}},\ \bibinfo {pages} {3257–3264} (\bibinfo {year} {2023})}\BibitemShut
  {NoStop}%
\bibitem [{\citenamefont {Sardar}\ \emph {et~al.}(2023)\citenamefont {Sardar},
  \citenamefont {Christianen}, \citenamefont {Li},\ and\ \citenamefont
  {Bohn}}]{SardarPhysRevA.107.2023}%
  \BibitemOpen
  \bibfield  {author} {\bibinfo {author} {\bibfnamefont {D.}~\bibnamefont
  {Sardar}}, \bibinfo {author} {\bibfnamefont {A.}~\bibnamefont {Christianen}},
  \bibinfo {author} {\bibfnamefont {H.}~\bibnamefont {Li}},\ and\ \bibinfo
  {author} {\bibfnamefont {J.~L.}\ \bibnamefont {Bohn}},\ }\bibfield  {title}
  {\bibinfo {title} {Four-body singlet potential-energy surface for reactions
  of calcium monofluoride},\ }\href
  {https://doi.org/10.1103/PhysRevA.107.032822} {\bibfield  {journal} {\bibinfo
   {journal} {Phys. Rev. A}\ }\textbf {\bibinfo {volume} {107}},\ \bibinfo
  {pages} {032822} (\bibinfo {year} {2023})}\BibitemShut {NoStop}%
\bibitem [{\citenamefont {K\l{}os}\ and\ \citenamefont
  {Kotochigova}(2020)}]{PhysRevResearch.2.013384}%
  \BibitemOpen
  \bibfield  {author} {\bibinfo {author} {\bibfnamefont {J.}~\bibnamefont
  {K\l{}os}}\ and\ \bibinfo {author} {\bibfnamefont {S.}~\bibnamefont
  {Kotochigova}},\ }\bibfield  {title} {\bibinfo {title} {Prospects for laser
  cooling of polyatomic molecules with increasing complexity},\ }\href
  {https://doi.org/10.1103/PhysRevResearch.2.013384} {\bibfield  {journal}
  {\bibinfo  {journal} {Phys. Rev. Res.}\ }\textbf {\bibinfo {volume} {2}},\
  \bibinfo {pages} {013384} (\bibinfo {year} {2020})}\BibitemShut {NoStop}%
\bibitem [{\citenamefont {Augenbraun}\ \emph
  {et~al.}(2020{\natexlab{a}})\citenamefont {Augenbraun}, \citenamefont
  {Doyle}, \citenamefont {Zelevinsky},\ and\ \citenamefont
  {Kozyryev}}]{PhysRevX.10.031022}%
  \BibitemOpen
  \bibfield  {author} {\bibinfo {author} {\bibfnamefont {B.~L.}\ \bibnamefont
  {Augenbraun}}, \bibinfo {author} {\bibfnamefont {J.~M.}\ \bibnamefont
  {Doyle}}, \bibinfo {author} {\bibfnamefont {T.}~\bibnamefont {Zelevinsky}},\
  and\ \bibinfo {author} {\bibfnamefont {I.}~\bibnamefont {Kozyryev}},\
  }\bibfield  {title} {\bibinfo {title} {Molecular asymmetry and optical
  cycling: Laser cooling asymmetric top molecules},\ }\href
  {https://doi.org/10.1103/PhysRevX.10.031022} {\bibfield  {journal} {\bibinfo
  {journal} {Phys. Rev. X}\ }\textbf {\bibinfo {volume} {10}},\ \bibinfo
  {pages} {031022} (\bibinfo {year} {2020}{\natexlab{a}})}\BibitemShut
  {NoStop}%
\bibitem [{\citenamefont {Prehn}\ \emph {et~al.}(2016)\citenamefont {Prehn},
  \citenamefont {Ibr{\"u}gger}, \citenamefont {Gl{\"o}ckner}, \citenamefont
  {Rempe},\ and\ \citenamefont {Zeppenfeld}}]{prehn2016Phys.Rev.Lett.}%
  \BibitemOpen
  \bibfield  {author} {\bibinfo {author} {\bibfnamefont {A.}~\bibnamefont
  {Prehn}}, \bibinfo {author} {\bibfnamefont {M.}~\bibnamefont {Ibr{\"u}gger}},
  \bibinfo {author} {\bibfnamefont {R.}~\bibnamefont {Gl{\"o}ckner}}, \bibinfo
  {author} {\bibfnamefont {G.}~\bibnamefont {Rempe}},\ and\ \bibinfo {author}
  {\bibfnamefont {M.}~\bibnamefont {Zeppenfeld}},\ }\bibfield  {title}
  {\bibinfo {title} {Optoelectrical {{Cooling}} of {{Polar Molecules}} to
  {{Submillikelvin Temperatures}}},\ }\href
  {https://doi.org/10.1103/PhysRevLett.116.063005} {\bibfield  {journal}
  {\bibinfo  {journal} {Phys. Rev. Lett.}\ }\textbf {\bibinfo {volume} {116}},\
  \bibinfo {pages} {063005} (\bibinfo {year} {2016})}\BibitemShut {NoStop}%
\bibitem [{\citenamefont {Kozyryev}\ \emph {et~al.}(2017)\citenamefont
  {Kozyryev}, \citenamefont {Baum}, \citenamefont {Matsuda}, \citenamefont
  {Augenbraun}, \citenamefont {Anderegg}, \citenamefont {Sedlack},\ and\
  \citenamefont {Doyle}}]{KozyryevPhysRevLett.118.2017}%
  \BibitemOpen
  \bibfield  {author} {\bibinfo {author} {\bibfnamefont {I.}~\bibnamefont
  {Kozyryev}}, \bibinfo {author} {\bibfnamefont {L.}~\bibnamefont {Baum}},
  \bibinfo {author} {\bibfnamefont {K.}~\bibnamefont {Matsuda}}, \bibinfo
  {author} {\bibfnamefont {B.~L.}\ \bibnamefont {Augenbraun}}, \bibinfo
  {author} {\bibfnamefont {L.}~\bibnamefont {Anderegg}}, \bibinfo {author}
  {\bibfnamefont {A.~P.}\ \bibnamefont {Sedlack}},\ and\ \bibinfo {author}
  {\bibfnamefont {J.~M.}\ \bibnamefont {Doyle}},\ }\bibfield  {title} {\bibinfo
  {title} {Sisyphus laser cooling of a polyatomic molecule},\ }\href
  {https://doi.org/10.1103/PhysRevLett.118.173201} {\bibfield  {journal}
  {\bibinfo  {journal} {Phys. Rev. Lett.}\ }\textbf {\bibinfo {volume} {118}},\
  \bibinfo {pages} {173201} (\bibinfo {year} {2017})}\BibitemShut {NoStop}%
\bibitem [{\citenamefont {Augenbraun}\ \emph
  {et~al.}(2020{\natexlab{b}})\citenamefont {Augenbraun}, \citenamefont
  {Lasner}, \citenamefont {Frenett}, \citenamefont {Sawaoka}, \citenamefont
  {Miller}, \citenamefont {Steimle},\ and\ \citenamefont
  {Doyle}}]{Augenbraun2020NJP}%
  \BibitemOpen
  \bibfield  {author} {\bibinfo {author} {\bibfnamefont {B.~L.}\ \bibnamefont
  {Augenbraun}}, \bibinfo {author} {\bibfnamefont {Z.~D.}\ \bibnamefont
  {Lasner}}, \bibinfo {author} {\bibfnamefont {A.}~\bibnamefont {Frenett}},
  \bibinfo {author} {\bibfnamefont {H.}~\bibnamefont {Sawaoka}}, \bibinfo
  {author} {\bibfnamefont {C.}~\bibnamefont {Miller}}, \bibinfo {author}
  {\bibfnamefont {T.~C.}\ \bibnamefont {Steimle}},\ and\ \bibinfo {author}
  {\bibfnamefont {J.~M.}\ \bibnamefont {Doyle}},\ }\bibfield  {title} {\bibinfo
  {title} {Laser-cooled polyatomic molecules for improved electron electric
  dipole moment searches},\ }\href {https://doi.org/10.1088/1367-2630/ab687b}
  {\bibfield  {journal} {\bibinfo  {journal} {New J. Phys.}\ }\textbf {\bibinfo
  {volume} {22}},\ \bibinfo {pages} {022003} (\bibinfo {year}
  {2020}{\natexlab{b}})}\BibitemShut {NoStop}%
\bibitem [{\citenamefont {Vilas}\ \emph {et~al.}(2022)\citenamefont {Vilas},
  \citenamefont {Hallas}, \citenamefont {Anderegg}, \citenamefont {Robichaud},
  \citenamefont {Winnicki}, \citenamefont {Mitra},\ and\ \citenamefont
  {Doyle}}]{vilas2022Nature}%
  \BibitemOpen
  \bibfield  {author} {\bibinfo {author} {\bibfnamefont {N.~B.}\ \bibnamefont
  {Vilas}}, \bibinfo {author} {\bibfnamefont {C.}~\bibnamefont {Hallas}},
  \bibinfo {author} {\bibfnamefont {L.}~\bibnamefont {Anderegg}}, \bibinfo
  {author} {\bibfnamefont {P.}~\bibnamefont {Robichaud}}, \bibinfo {author}
  {\bibfnamefont {A.}~\bibnamefont {Winnicki}}, \bibinfo {author}
  {\bibfnamefont {D.}~\bibnamefont {Mitra}},\ and\ \bibinfo {author}
  {\bibfnamefont {J.~M.}\ \bibnamefont {Doyle}},\ }\bibfield  {title} {\bibinfo
  {title} {Magneto-optical trapping and sub-{{Doppler}} cooling of a polyatomic
  molecule},\ }\href {https://doi.org/10.1038/s41586-022-04620-5} {\bibfield
  {journal} {\bibinfo  {journal} {Nature}\ }\textbf {\bibinfo {volume} {606}},\
  \bibinfo {pages} {70} (\bibinfo {year} {2022})}\BibitemShut {NoStop}%
\bibitem [{\citenamefont {Mitra}\ \emph {et~al.}(2020)\citenamefont {Mitra},
  \citenamefont {Vilas}, \citenamefont {Hallas}, \citenamefont {Anderegg},
  \citenamefont {Augenbraun}, \citenamefont {Baum}, \citenamefont {Miller},
  \citenamefont {Raval},\ and\ \citenamefont {Doyle}}]{mitra2020Science}%
  \BibitemOpen
  \bibfield  {author} {\bibinfo {author} {\bibfnamefont {D.}~\bibnamefont
  {Mitra}}, \bibinfo {author} {\bibfnamefont {N.~B.}\ \bibnamefont {Vilas}},
  \bibinfo {author} {\bibfnamefont {C.}~\bibnamefont {Hallas}}, \bibinfo
  {author} {\bibfnamefont {L.}~\bibnamefont {Anderegg}}, \bibinfo {author}
  {\bibfnamefont {B.~L.}\ \bibnamefont {Augenbraun}}, \bibinfo {author}
  {\bibfnamefont {L.}~\bibnamefont {Baum}}, \bibinfo {author} {\bibfnamefont
  {C.}~\bibnamefont {Miller}}, \bibinfo {author} {\bibfnamefont
  {S.}~\bibnamefont {Raval}},\ and\ \bibinfo {author} {\bibfnamefont {J.~M.}\
  \bibnamefont {Doyle}},\ }\bibfield  {title} {\bibinfo {title} {Direct laser
  cooling of a symmetric top molecule},\ }\href
  {https://doi.org/10.1126/science.abc5357} {\bibfield  {journal} {\bibinfo
  {journal} {Science}\ }\textbf {\bibinfo {volume} {369}},\ \bibinfo {pages}
  {1366} (\bibinfo {year} {2020})}\BibitemShut {NoStop}%
\bibitem [{\citenamefont {Langen}\ \emph {et~al.}(2024)\citenamefont {Langen},
  \citenamefont {Valtolina}, \citenamefont {Wang},\ and\ \citenamefont
  {Ye}}]{LangenNP2024}%
  \BibitemOpen
  \bibfield  {author} {\bibinfo {author} {\bibfnamefont {T.}~\bibnamefont
  {Langen}}, \bibinfo {author} {\bibfnamefont {G.}~\bibnamefont {Valtolina}},
  \bibinfo {author} {\bibfnamefont {D.}~\bibnamefont {Wang}},\ and\ \bibinfo
  {author} {\bibfnamefont {J.}~\bibnamefont {Ye}},\ }\bibfield  {title}
  {\bibinfo {title} {Quantum state manipulation and cooling of ultracold
  molecules},\ }\href {https://doi.org/10.1038/s41567-024-02423-1} {\bibfield
  {journal} {\bibinfo  {journal} {Nat. Phys.}\ }\textbf {\bibinfo {volume}
  {20}},\ \bibinfo {pages} {702–712} (\bibinfo {year} {2024})}\BibitemShut
  {NoStop}%
\bibitem [{\citenamefont {Yang}\ \emph {et~al.}(2019)\citenamefont {Yang},
  \citenamefont {Zhang}, \citenamefont {Liu}, \citenamefont {Liu},
  \citenamefont {Nan}, \citenamefont {Zhao},\ and\ \citenamefont
  {Pan}}]{yang2019Science}%
  \BibitemOpen
  \bibfield  {author} {\bibinfo {author} {\bibfnamefont {H.}~\bibnamefont
  {Yang}}, \bibinfo {author} {\bibfnamefont {D.-C.}\ \bibnamefont {Zhang}},
  \bibinfo {author} {\bibfnamefont {L.}~\bibnamefont {Liu}}, \bibinfo {author}
  {\bibfnamefont {Y.-X.}\ \bibnamefont {Liu}}, \bibinfo {author} {\bibfnamefont
  {J.}~\bibnamefont {Nan}}, \bibinfo {author} {\bibfnamefont {B.}~\bibnamefont
  {Zhao}},\ and\ \bibinfo {author} {\bibfnamefont {J.-W.}\ \bibnamefont
  {Pan}},\ }\bibfield  {title} {\bibinfo {title} {Observation of magnetically
  tunable {{Feshbach}} resonances in ultracold {\textsuperscript{23}} {{Na}}
  {\textsuperscript{40}} {{K}} + {\textsuperscript{40}} {{K}} collisions},\
  }\href {https://doi.org/10.1126/science.aau5322} {\bibfield  {journal}
  {\bibinfo  {journal} {Science}\ }\textbf {\bibinfo {volume} {363}},\ \bibinfo
  {pages} {261} (\bibinfo {year} {2019})}\BibitemShut {NoStop}%
\bibitem [{\citenamefont {Chen}\ \emph {et~al.}(2023)\citenamefont {Chen},
  \citenamefont {Schindewolf}, \citenamefont {Eppelt}, \citenamefont {Bause},
  \citenamefont {Duda}, \citenamefont {Biswas}, \citenamefont {Karman},
  \citenamefont {Hilker}, \citenamefont {Bloch},\ and\ \citenamefont
  {Luo}}]{chen2023Nature}%
  \BibitemOpen
  \bibfield  {author} {\bibinfo {author} {\bibfnamefont {X.-Y.}\ \bibnamefont
  {Chen}}, \bibinfo {author} {\bibfnamefont {A.}~\bibnamefont {Schindewolf}},
  \bibinfo {author} {\bibfnamefont {S.}~\bibnamefont {Eppelt}}, \bibinfo
  {author} {\bibfnamefont {R.}~\bibnamefont {Bause}}, \bibinfo {author}
  {\bibfnamefont {M.}~\bibnamefont {Duda}}, \bibinfo {author} {\bibfnamefont
  {S.}~\bibnamefont {Biswas}}, \bibinfo {author} {\bibfnamefont
  {T.}~\bibnamefont {Karman}}, \bibinfo {author} {\bibfnamefont
  {T.}~\bibnamefont {Hilker}}, \bibinfo {author} {\bibfnamefont
  {I.}~\bibnamefont {Bloch}},\ and\ \bibinfo {author} {\bibfnamefont {X.-Y.}\
  \bibnamefont {Luo}},\ }\bibfield  {title} {\bibinfo {title} {Field-linked
  resonances of polar molecules},\ }\href
  {https://doi.org/10.1038/s41586-022-05651-8} {\bibfield  {journal} {\bibinfo
  {journal} {Nature}\ }\textbf {\bibinfo {volume} {614}},\ \bibinfo {pages}
  {59} (\bibinfo {year} {2023})}\BibitemShut {NoStop}%
\bibitem [{\citenamefont {Li}\ \emph {et~al.}(2023)\citenamefont {Li},
  \citenamefont {Gu}, \citenamefont {Wang},\ and\ \citenamefont
  {Zhang}}]{li2023Sci.ChinaPhys.Mech.Astron.}%
  \BibitemOpen
  \bibfield  {author} {\bibinfo {author} {\bibfnamefont {Z.-L.}\ \bibnamefont
  {Li}}, \bibinfo {author} {\bibfnamefont {Z.-Y.}\ \bibnamefont {Gu}}, \bibinfo
  {author} {\bibfnamefont {P.-J.}\ \bibnamefont {Wang}},\ and\ \bibinfo
  {author} {\bibfnamefont {J.}~\bibnamefont {Zhang}},\ }\bibfield  {title}
  {\bibinfo {title} {Efficient production of ultracold polar molecules
  {{23Na40K}} in their absolute ground state via intermediate state of the
  coupled complex {{B1$\Pi$}}{\textbar}{$\nu$} = 4{\textrangle} {$\sim$}
  {{c3$\Sigma$}}+ {\textbar}{$\nu$} = 25{\textrangle}},\ }\href
  {https://doi.org/10.1007/s11433-023-2148-8} {\bibfield  {journal} {\bibinfo
  {journal} {Sci. China Phys. Mech. Astron.}\ }\textbf {\bibinfo {volume}
  {66}},\ \bibinfo {pages} {293011} (\bibinfo {year} {2023})}\BibitemShut
  {NoStop}%
\bibitem [{\citenamefont {Will}\ \emph {et~al.}(2016)\citenamefont {Will},
  \citenamefont {Park}, \citenamefont {Yan}, \citenamefont {Loh},\ and\
  \citenamefont {Zwierlein}}]{will2016Phys.Rev.Lett.}%
  \BibitemOpen
  \bibfield  {author} {\bibinfo {author} {\bibfnamefont {S.~A.}\ \bibnamefont
  {Will}}, \bibinfo {author} {\bibfnamefont {J.~W.}\ \bibnamefont {Park}},
  \bibinfo {author} {\bibfnamefont {Z.~Z.}\ \bibnamefont {Yan}}, \bibinfo
  {author} {\bibfnamefont {H.}~\bibnamefont {Loh}},\ and\ \bibinfo {author}
  {\bibfnamefont {M.~W.}\ \bibnamefont {Zwierlein}},\ }\bibfield  {title}
  {\bibinfo {title} {Coherent {{Microwave Control}} of {{Ultracold Na}} 23
  {{K}} 40 {{Molecules}}},\ }\href
  {https://doi.org/10.1103/PhysRevLett.116.225306} {\bibfield  {journal}
  {\bibinfo  {journal} {Phys. Rev. Lett.}\ }\textbf {\bibinfo {volume} {116}},\
  \bibinfo {pages} {225306} (\bibinfo {year} {2016})}\BibitemShut {NoStop}%
\bibitem [{\citenamefont {Cohen-Tannoudji}\ \emph {et~al.}(1998)\citenamefont
  {Cohen-Tannoudji}, \citenamefont {Dupont-Roc},\ and\ \citenamefont
  {Grynberg}}]{CohenTannoudji1998}%
  \BibitemOpen
  \bibfield  {author} {\bibinfo {author} {\bibfnamefont {C.}~\bibnamefont
  {Cohen-Tannoudji}}, \bibinfo {author} {\bibfnamefont {J.}~\bibnamefont
  {Dupont-Roc}},\ and\ \bibinfo {author} {\bibfnamefont {G.}~\bibnamefont
  {Grynberg}},\ }\href@noop {} {\emph {\bibinfo {title} {Atom-Photon
  Interactions: Basic Processes and Applications}}}\ (\bibinfo  {publisher}
  {John Wiley \& Sons},\ \bibinfo {address} {New York},\ \bibinfo {year}
  {1998})\BibitemShut {NoStop}%
\bibitem [{\citenamefont {Zhang}\ \emph {et~al.}(2024)\citenamefont {Zhang},
  \citenamefont {Yuan}, \citenamefont {Bigagli}, \citenamefont {Warner},
  \citenamefont {Stevenson},\ and\ \citenamefont
  {Will}}]{zhang2024Phys.Rev.Lett.}%
  \BibitemOpen
  \bibfield  {author} {\bibinfo {author} {\bibfnamefont {S.}~\bibnamefont
  {Zhang}}, \bibinfo {author} {\bibfnamefont {W.}~\bibnamefont {Yuan}},
  \bibinfo {author} {\bibfnamefont {N.}~\bibnamefont {Bigagli}}, \bibinfo
  {author} {\bibfnamefont {C.}~\bibnamefont {Warner}}, \bibinfo {author}
  {\bibfnamefont {I.}~\bibnamefont {Stevenson}},\ and\ \bibinfo {author}
  {\bibfnamefont {S.}~\bibnamefont {Will}},\ }\bibfield  {title} {\bibinfo
  {title} {Dressed-state spectroscopy and magic trapping of microwave-shielded
  {{NaCs}} molecules},\ }\href {https://doi.org/10.1103/PhysRevLett.133.263401}
  {\bibfield  {journal} {\bibinfo  {journal} {Physical Review Letters}\
  }\textbf {\bibinfo {volume} {133}},\ \bibinfo {pages} {263401} (\bibinfo
  {year} {2024})}\BibitemShut {NoStop}%
\bibitem [{\citenamefont {Ospelkaus}\ \emph {et~al.}(2006)\citenamefont
  {Ospelkaus}, \citenamefont {Ospelkaus}, \citenamefont {Humbert},
  \citenamefont {Ernst}, \citenamefont {Sengstock},\ and\ \citenamefont
  {Bongs}}]{PhysRevLett.97.120402}%
  \BibitemOpen
  \bibfield  {author} {\bibinfo {author} {\bibfnamefont {C.}~\bibnamefont
  {Ospelkaus}}, \bibinfo {author} {\bibfnamefont {S.}~\bibnamefont
  {Ospelkaus}}, \bibinfo {author} {\bibfnamefont {L.}~\bibnamefont {Humbert}},
  \bibinfo {author} {\bibfnamefont {P.}~\bibnamefont {Ernst}}, \bibinfo
  {author} {\bibfnamefont {K.}~\bibnamefont {Sengstock}},\ and\ \bibinfo
  {author} {\bibfnamefont {K.}~\bibnamefont {Bongs}},\ }\bibfield  {title}
  {\bibinfo {title} {Ultracold heteronuclear molecules in a 3d optical
  lattice},\ }\href {https://doi.org/10.1103/PhysRevLett.97.120402} {\bibfield
  {journal} {\bibinfo  {journal} {Phys. Rev. Lett.}\ }\textbf {\bibinfo
  {volume} {97}},\ \bibinfo {pages} {120402} (\bibinfo {year}
  {2006})}\BibitemShut {NoStop}%
\bibitem [{\citenamefont {Zirbel}\ \emph {et~al.}(2008)\citenamefont {Zirbel},
  \citenamefont {Ni}, \citenamefont {Ospelkaus}, \citenamefont {D'Incao},
  \citenamefont {Wieman}, \citenamefont {Ye},\ and\ \citenamefont
  {Jin}}]{PhysRevLett.100.143201}%
  \BibitemOpen
  \bibfield  {author} {\bibinfo {author} {\bibfnamefont {J.~J.}\ \bibnamefont
  {Zirbel}}, \bibinfo {author} {\bibfnamefont {K.-K.}\ \bibnamefont {Ni}},
  \bibinfo {author} {\bibfnamefont {S.}~\bibnamefont {Ospelkaus}}, \bibinfo
  {author} {\bibfnamefont {J.~P.}\ \bibnamefont {D'Incao}}, \bibinfo {author}
  {\bibfnamefont {C.~E.}\ \bibnamefont {Wieman}}, \bibinfo {author}
  {\bibfnamefont {J.}~\bibnamefont {Ye}},\ and\ \bibinfo {author}
  {\bibfnamefont {D.~S.}\ \bibnamefont {Jin}},\ }\bibfield  {title} {\bibinfo
  {title} {Collisional stability of fermionic feshbach molecules},\ }\href
  {https://doi.org/10.1103/PhysRevLett.100.143201} {\bibfield  {journal}
  {\bibinfo  {journal} {Phys. Rev. Lett.}\ }\textbf {\bibinfo {volume} {100}},\
  \bibinfo {pages} {143201} (\bibinfo {year} {2008})}\BibitemShut {NoStop}%
\bibitem [{\citenamefont {Fu}\ \emph {et~al.}(2013)\citenamefont {Fu},
  \citenamefont {Huang}, \citenamefont {Meng}, \citenamefont {Wang},
  \citenamefont {Zhang}, \citenamefont {Zhang}, \citenamefont {Zhai},
  \citenamefont {Zhang},\ and\ \citenamefont {Zhang}}]{Fu2013}%
  \BibitemOpen
  \bibfield  {author} {\bibinfo {author} {\bibfnamefont {Z.}~\bibnamefont
  {Fu}}, \bibinfo {author} {\bibfnamefont {L.}~\bibnamefont {Huang}}, \bibinfo
  {author} {\bibfnamefont {Z.}~\bibnamefont {Meng}}, \bibinfo {author}
  {\bibfnamefont {P.}~\bibnamefont {Wang}}, \bibinfo {author} {\bibfnamefont
  {L.}~\bibnamefont {Zhang}}, \bibinfo {author} {\bibfnamefont
  {S.}~\bibnamefont {Zhang}}, \bibinfo {author} {\bibfnamefont
  {H.}~\bibnamefont {Zhai}}, \bibinfo {author} {\bibfnamefont {P.}~\bibnamefont
  {Zhang}},\ and\ \bibinfo {author} {\bibfnamefont {J.}~\bibnamefont {Zhang}},\
  }\bibfield  {title} {\bibinfo {title} {Production of feshbach molecules
  induced by spin–orbit coupling in fermi gases},\ }\href
  {https://doi.org/10.1038/nphys2824} {\bibfield  {journal} {\bibinfo
  {journal} {Nature Physics}\ }\textbf {\bibinfo {volume} {10}},\ \bibinfo
  {pages} {110–115} (\bibinfo {year} {2013})}\BibitemShut {NoStop}%
\bibitem [{\citenamefont {Lassabli\`ere}\ and\ \citenamefont
  {Qu\'em\'ener}(2018)}]{PhysRevLett.121.163402}%
  \BibitemOpen
  \bibfield  {author} {\bibinfo {author} {\bibfnamefont {L.}~\bibnamefont
  {Lassabli\`ere}}\ and\ \bibinfo {author} {\bibfnamefont {G.}~\bibnamefont
  {Qu\'em\'ener}},\ }\bibfield  {title} {\bibinfo {title} {Controlling the
  scattering length of ultracold dipolar molecules},\ }\href
  {https://doi.org/10.1103/PhysRevLett.121.163402} {\bibfield  {journal}
  {\bibinfo  {journal} {Phys. Rev. Lett.}\ }\textbf {\bibinfo {volume} {121}},\
  \bibinfo {pages} {163402} (\bibinfo {year} {2018})}\BibitemShut {NoStop}%
\bibitem [{\citenamefont {Fedichev}\ \emph {et~al.}(1996)\citenamefont
  {Fedichev}, \citenamefont {Kagan}, \citenamefont {Shlyapnikov},\ and\
  \citenamefont {Walraven}}]{PhysRevLett.77.2913}%
  \BibitemOpen
  \bibfield  {author} {\bibinfo {author} {\bibfnamefont {P.~O.}\ \bibnamefont
  {Fedichev}}, \bibinfo {author} {\bibfnamefont {Y.}~\bibnamefont {Kagan}},
  \bibinfo {author} {\bibfnamefont {G.~V.}\ \bibnamefont {Shlyapnikov}},\ and\
  \bibinfo {author} {\bibfnamefont {J.~T.~M.}\ \bibnamefont {Walraven}},\
  }\bibfield  {title} {\bibinfo {title} {Influence of nearly resonant light on
  the scattering length in low-temperature atomic gases},\ }\href
  {https://doi.org/10.1103/PhysRevLett.77.2913} {\bibfield  {journal} {\bibinfo
   {journal} {Phys. Rev. Lett.}\ }\textbf {\bibinfo {volume} {77}},\ \bibinfo
  {pages} {2913} (\bibinfo {year} {1996})}\BibitemShut {NoStop}%
\bibitem [{\citenamefont {Bohn}\ and\ \citenamefont
  {Julienne}(1997)}]{PhysRevA.56.1486}%
  \BibitemOpen
  \bibfield  {author} {\bibinfo {author} {\bibfnamefont {J.~L.}\ \bibnamefont
  {Bohn}}\ and\ \bibinfo {author} {\bibfnamefont {P.~S.}\ \bibnamefont
  {Julienne}},\ }\bibfield  {title} {\bibinfo {title} {Prospects for
  influencing scattering lengths with far-off-resonant light},\ }\href
  {https://doi.org/10.1103/PhysRevA.56.1486} {\bibfield  {journal} {\bibinfo
  {journal} {Phys. Rev. A}\ }\textbf {\bibinfo {volume} {56}},\ \bibinfo
  {pages} {1486} (\bibinfo {year} {1997})}\BibitemShut {NoStop}%
\bibitem [{\citenamefont {Fatemi}\ \emph {et~al.}(2000)\citenamefont {Fatemi},
  \citenamefont {Jones},\ and\ \citenamefont {Lett}}]{PhysRevLett.85.4462}%
  \BibitemOpen
  \bibfield  {author} {\bibinfo {author} {\bibfnamefont {F.~K.}\ \bibnamefont
  {Fatemi}}, \bibinfo {author} {\bibfnamefont {K.~M.}\ \bibnamefont {Jones}},\
  and\ \bibinfo {author} {\bibfnamefont {P.~D.}\ \bibnamefont {Lett}},\
  }\bibfield  {title} {\bibinfo {title} {Observation of optically induced
  feshbach resonances in collisions of cold atoms},\ }\href
  {https://doi.org/10.1103/PhysRevLett.85.4462} {\bibfield  {journal} {\bibinfo
   {journal} {Phys. Rev. Lett.}\ }\textbf {\bibinfo {volume} {85}},\ \bibinfo
  {pages} {4462} (\bibinfo {year} {2000})}\BibitemShut {NoStop}%
\bibitem [{\citenamefont {Schindewolf}\ \emph {et~al.}(2022)\citenamefont
  {Schindewolf}, \citenamefont {Bause}, \citenamefont {Chen}, \citenamefont
  {Duda}, \citenamefont {Karman}, \citenamefont {Bloch},\ and\ \citenamefont
  {Luo}}]{schindewolf2022Nature}%
  \BibitemOpen
  \bibfield  {author} {\bibinfo {author} {\bibfnamefont {A.}~\bibnamefont
  {Schindewolf}}, \bibinfo {author} {\bibfnamefont {R.}~\bibnamefont {Bause}},
  \bibinfo {author} {\bibfnamefont {X.-Y.}\ \bibnamefont {Chen}}, \bibinfo
  {author} {\bibfnamefont {M.}~\bibnamefont {Duda}}, \bibinfo {author}
  {\bibfnamefont {T.}~\bibnamefont {Karman}}, \bibinfo {author} {\bibfnamefont
  {I.}~\bibnamefont {Bloch}},\ and\ \bibinfo {author} {\bibfnamefont {X.-Y.}\
  \bibnamefont {Luo}},\ }\bibfield  {title} {\bibinfo {title} {Evaporation of
  microwave-shielded polar molecules to quantum degeneracy},\ }\href
  {https://doi.org/10.1038/s41586-022-04900-0} {\bibfield  {journal} {\bibinfo
  {journal} {Nature}\ }\textbf {\bibinfo {volume} {607}},\ \bibinfo {pages}
  {677} (\bibinfo {year} {2022})}\BibitemShut {NoStop}%
\bibitem [{\citenamefont {Bigagli}\ \emph {et~al.}(2023)\citenamefont
  {Bigagli}, \citenamefont {Warner}, \citenamefont {Yuan}, \citenamefont
  {Zhang}, \citenamefont {Stevenson}, \citenamefont {Karman},\ and\
  \citenamefont {Will}}]{Bigagli2023Nat.Phys.}%
  \BibitemOpen
  \bibfield  {author} {\bibinfo {author} {\bibfnamefont {N.}~\bibnamefont
  {Bigagli}}, \bibinfo {author} {\bibfnamefont {C.}~\bibnamefont {Warner}},
  \bibinfo {author} {\bibfnamefont {W.}~\bibnamefont {Yuan}}, \bibinfo {author}
  {\bibfnamefont {S.}~\bibnamefont {Zhang}}, \bibinfo {author} {\bibfnamefont
  {I.}~\bibnamefont {Stevenson}}, \bibinfo {author} {\bibfnamefont
  {T.}~\bibnamefont {Karman}},\ and\ \bibinfo {author} {\bibfnamefont
  {S.}~\bibnamefont {Will}},\ }\bibfield  {title} {\bibinfo {title}
  {Collisionally stable gas of bosonic dipolar ground-state molecules},\ }\href
  {https://doi.org/10.1038/s41567-023-02200-6} {\bibfield  {journal} {\bibinfo
  {journal} {Nat. Phys.}\ }\textbf {\bibinfo {volume} {19}},\ \bibinfo {pages}
  {1579–1584} (\bibinfo {year} {2023})}\BibitemShut {NoStop}%
\bibitem [{\citenamefont {Lin}\ \emph {et~al.}(2023)\citenamefont {Lin},
  \citenamefont {Chen}, \citenamefont {Jin}, \citenamefont {Shi}, \citenamefont
  {Deng}, \citenamefont {Zhang}, \citenamefont {Qu{\'e}m{\'e}ner},
  \citenamefont {Shi}, \citenamefont {Yi},\ and\ \citenamefont
  {Wang}}]{lin2023Phys.Rev.X}%
  \BibitemOpen
  \bibfield  {author} {\bibinfo {author} {\bibfnamefont {J.}~\bibnamefont
  {Lin}}, \bibinfo {author} {\bibfnamefont {G.}~\bibnamefont {Chen}}, \bibinfo
  {author} {\bibfnamefont {M.}~\bibnamefont {Jin}}, \bibinfo {author}
  {\bibfnamefont {Z.}~\bibnamefont {Shi}}, \bibinfo {author} {\bibfnamefont
  {F.}~\bibnamefont {Deng}}, \bibinfo {author} {\bibfnamefont {W.}~\bibnamefont
  {Zhang}}, \bibinfo {author} {\bibfnamefont {G.}~\bibnamefont
  {Qu{\'e}m{\'e}ner}}, \bibinfo {author} {\bibfnamefont {T.}~\bibnamefont
  {Shi}}, \bibinfo {author} {\bibfnamefont {S.}~\bibnamefont {Yi}},\ and\
  \bibinfo {author} {\bibfnamefont {D.}~\bibnamefont {Wang}},\ }\bibfield
  {title} {\bibinfo {title} {Microwave {{Shielding}} of {{Bosonic NaRb
  Molecules}}},\ }\href {https://doi.org/10.1103/PhysRevX.13.031032} {\bibfield
   {journal} {\bibinfo  {journal} {Phys. Rev. X}\ }\textbf {\bibinfo {volume}
  {13}},\ \bibinfo {pages} {031032} (\bibinfo {year} {2023})}\BibitemShut
  {NoStop}%
\bibitem [{\citenamefont {Anderegg}\ \emph {et~al.}(2021)\citenamefont
  {Anderegg}, \citenamefont {Burchesky}, \citenamefont {Bao}, \citenamefont
  {Yu}, \citenamefont {Karman}, \citenamefont {Chae}, \citenamefont {Ni},
  \citenamefont {Ketterle},\ and\ \citenamefont {Doyle}}]{anderegg2021Science}%
  \BibitemOpen
  \bibfield  {author} {\bibinfo {author} {\bibfnamefont {L.}~\bibnamefont
  {Anderegg}}, \bibinfo {author} {\bibfnamefont {S.}~\bibnamefont {Burchesky}},
  \bibinfo {author} {\bibfnamefont {Y.}~\bibnamefont {Bao}}, \bibinfo {author}
  {\bibfnamefont {S.~S.}\ \bibnamefont {Yu}}, \bibinfo {author} {\bibfnamefont
  {T.}~\bibnamefont {Karman}}, \bibinfo {author} {\bibfnamefont
  {E.}~\bibnamefont {Chae}}, \bibinfo {author} {\bibfnamefont {K.-K.}\
  \bibnamefont {Ni}}, \bibinfo {author} {\bibfnamefont {W.}~\bibnamefont
  {Ketterle}},\ and\ \bibinfo {author} {\bibfnamefont {J.~M.}\ \bibnamefont
  {Doyle}},\ }\bibfield  {title} {\bibinfo {title} {Observation of microwave
  shielding of ultracold molecules},\ }\href
  {https://doi.org/10.1126/science.abg9502} {\bibfield  {journal} {\bibinfo
  {journal} {Science}\ }\textbf {\bibinfo {volume} {373}},\ \bibinfo {pages}
  {779} (\bibinfo {year} {2021})}\BibitemShut {NoStop}%
\bibitem [{\citenamefont {Bigagli}\ \emph {et~al.}(2024)\citenamefont
  {Bigagli}, \citenamefont {Yuan}, \citenamefont {Zhang}, \citenamefont
  {Bulatovic}, \citenamefont {Karman}, \citenamefont {Stevenson},\ and\
  \citenamefont {Will}}]{bigagli2024Naturea}%
  \BibitemOpen
  \bibfield  {author} {\bibinfo {author} {\bibfnamefont {N.}~\bibnamefont
  {Bigagli}}, \bibinfo {author} {\bibfnamefont {W.}~\bibnamefont {Yuan}},
  \bibinfo {author} {\bibfnamefont {S.}~\bibnamefont {Zhang}}, \bibinfo
  {author} {\bibfnamefont {B.}~\bibnamefont {Bulatovic}}, \bibinfo {author}
  {\bibfnamefont {T.}~\bibnamefont {Karman}}, \bibinfo {author} {\bibfnamefont
  {I.}~\bibnamefont {Stevenson}},\ and\ \bibinfo {author} {\bibfnamefont
  {S.}~\bibnamefont {Will}},\ }\bibfield  {title} {\bibinfo {title}
  {Observation of {{Bose}}--{{Einstein}} condensation of dipolar molecules},\
  }\href {https://doi.org/10.1038/s41586-024-07492-z} {\bibfield  {journal}
  {\bibinfo  {journal} {Nature}\ }\textbf {\bibinfo {volume} {631}},\ \bibinfo
  {pages} {289} (\bibinfo {year} {2024})}\BibitemShut {NoStop}%
\bibitem [{\citenamefont {Shi}\ \emph {et~al.}(2025)\citenamefont {Shi},
  \citenamefont {Huang}, \citenamefont {Deng}, \citenamefont {Jin},
  \citenamefont {Yi}, \citenamefont {Shi},\ and\ \citenamefont
  {Wang}}]{shi2025}%
  \BibitemOpen
  \bibfield  {author} {\bibinfo {author} {\bibfnamefont {Z.}~\bibnamefont
  {Shi}}, \bibinfo {author} {\bibfnamefont {Z.}~\bibnamefont {Huang}}, \bibinfo
  {author} {\bibfnamefont {F.}~\bibnamefont {Deng}}, \bibinfo {author}
  {\bibfnamefont {W.-J.}\ \bibnamefont {Jin}}, \bibinfo {author} {\bibfnamefont
  {S.}~\bibnamefont {Yi}}, \bibinfo {author} {\bibfnamefont {T.}~\bibnamefont
  {Shi}},\ and\ \bibinfo {author} {\bibfnamefont {D.}~\bibnamefont {Wang}},\
  }\href@noop {} {\bibinfo {title} {Bose-{{Einstein}} condensate of ultracold
  sodium-rubidium molecules with tunable dipolar interactions}} (\bibinfo
  {year} {2025}),\ \Eprint {https://arxiv.org/abs/2508.20518}
  {arXiv:2508.20518} \BibitemShut {NoStop}%
\bibitem [{\citenamefont {Park}\ \emph {et~al.}(2015)\citenamefont {Park},
  \citenamefont {Will},\ and\ \citenamefont
  {Zwierlein}}]{park2015Phys.Rev.Lett.}%
  \BibitemOpen
  \bibfield  {author} {\bibinfo {author} {\bibfnamefont {J.~W.}\ \bibnamefont
  {Park}}, \bibinfo {author} {\bibfnamefont {S.~A.}\ \bibnamefont {Will}},\
  and\ \bibinfo {author} {\bibfnamefont {M.~W.}\ \bibnamefont {Zwierlein}},\
  }\bibfield  {title} {\bibinfo {title} {Ultracold {{Dipolar Gas}} of
  {{Fermionic Na}} 23 {{K}} 40 {{Molecules}} in {{Their Absolute Ground
  State}}},\ }\href {https://doi.org/10.1103/PhysRevLett.114.205302} {\bibfield
   {journal} {\bibinfo  {journal} {Phys. Rev. Lett.}\ }\textbf {\bibinfo
  {volume} {114}},\ \bibinfo {pages} {205302} (\bibinfo {year}
  {2015})}\BibitemShut {NoStop}%
\bibitem [{\citenamefont {Bause}\ \emph {et~al.}(2021)\citenamefont {Bause},
  \citenamefont {Kamijo}, \citenamefont {Chen}, \citenamefont {Duda},
  \citenamefont {Schindewolf}, \citenamefont {Bloch},\ and\ \citenamefont
  {Luo}}]{bause2021Phys.Rev.A}%
  \BibitemOpen
  \bibfield  {author} {\bibinfo {author} {\bibfnamefont {R.}~\bibnamefont
  {Bause}}, \bibinfo {author} {\bibfnamefont {A.}~\bibnamefont {Kamijo}},
  \bibinfo {author} {\bibfnamefont {X.-Y.}\ \bibnamefont {Chen}}, \bibinfo
  {author} {\bibfnamefont {M.}~\bibnamefont {Duda}}, \bibinfo {author}
  {\bibfnamefont {A.}~\bibnamefont {Schindewolf}}, \bibinfo {author}
  {\bibfnamefont {I.}~\bibnamefont {Bloch}},\ and\ \bibinfo {author}
  {\bibfnamefont {X.-Y.}\ \bibnamefont {Luo}},\ }\bibfield  {title} {\bibinfo
  {title} {Efficient conversion of closed-channel-dominated {{Feshbach}}
  molecules of {{Na}} 23 {{K}} 40 to their absolute ground state},\ }\href
  {https://doi.org/10.1103/PhysRevA.104.043321} {\bibfield  {journal} {\bibinfo
   {journal} {Phys. Rev. A}\ }\textbf {\bibinfo {volume} {104}},\ \bibinfo
  {pages} {043321} (\bibinfo {year} {2021})}\BibitemShut {NoStop}%
\bibitem [{\citenamefont {Blackmore}\ \emph {et~al.}(2023)\citenamefont
  {Blackmore}, \citenamefont {Gregory}, \citenamefont {Hutson},\ and\
  \citenamefont {Cornish}}]{blackmore2023ComputerPhysicsCommunications}%
  \BibitemOpen
  \bibfield  {author} {\bibinfo {author} {\bibfnamefont {J.~A.}\ \bibnamefont
  {Blackmore}}, \bibinfo {author} {\bibfnamefont {P.~D.}\ \bibnamefont
  {Gregory}}, \bibinfo {author} {\bibfnamefont {J.~M.}\ \bibnamefont
  {Hutson}},\ and\ \bibinfo {author} {\bibfnamefont {S.~L.}\ \bibnamefont
  {Cornish}},\ }\bibfield  {title} {\bibinfo {title} {Diatomic-py: {{A Python}}
  module for calculating the rotational and hyperfine structure of
  {{$^{1}$$\Sigma$}} molecules},\ }\href
  {https://doi.org/10.1016/j.cpc.2022.108512} {\bibfield  {journal} {\bibinfo
  {journal} {Comput. Phys. Commun.}\ }\textbf {\bibinfo {volume} {282}},\
  \bibinfo {pages} {108512} (\bibinfo {year} {2023})}\BibitemShut {NoStop}%
\bibitem [{\citenamefont {Deng}\ \emph {et~al.}(2024)\citenamefont {Deng},
  \citenamefont {Chen}, \citenamefont {Luo}, \citenamefont {Zhang},
  \citenamefont {Yi},\ and\ \citenamefont {Shi}}]{deng2024}%
  \BibitemOpen
  \bibfield  {author} {\bibinfo {author} {\bibfnamefont {F.}~\bibnamefont
  {Deng}}, \bibinfo {author} {\bibfnamefont {X.-Y.}\ \bibnamefont {Chen}},
  \bibinfo {author} {\bibfnamefont {X.-Y.}\ \bibnamefont {Luo}}, \bibinfo
  {author} {\bibfnamefont {W.}~\bibnamefont {Zhang}}, \bibinfo {author}
  {\bibfnamefont {S.}~\bibnamefont {Yi}},\ and\ \bibinfo {author}
  {\bibfnamefont {T.}~\bibnamefont {Shi}},\ }\href@noop {} {\bibinfo {title}
  {Formation and {{Dissociation}} of {{Field-Linked Tetramers}}}} (\bibinfo
  {year} {2024}),\ \Eprint {https://arxiv.org/abs/2405.13645} {arXiv:2405.13645
  [quant-ph]} \BibitemShut {NoStop}%
\bibitem [{\citenamefont {Idziaszek}\ and\ \citenamefont
  {Julienne}(2010)}]{PhysRevLett.104.2010}%
  \BibitemOpen
  \bibfield  {author} {\bibinfo {author} {\bibfnamefont {Z.}~\bibnamefont
  {Idziaszek}}\ and\ \bibinfo {author} {\bibfnamefont {P.~S.}\ \bibnamefont
  {Julienne}},\ }\bibfield  {title} {\bibinfo {title} {Universal rate constants
  for reactive collisions of ultracold molecules},\ }\href
  {https://doi.org/10.1103/PhysRevLett.104.113202} {\bibfield  {journal}
  {\bibinfo  {journal} {Phys. Rev. Lett.}\ }\textbf {\bibinfo {volume} {104}},\
  \bibinfo {pages} {113202} (\bibinfo {year} {2010})}\BibitemShut {NoStop}%
\bibitem [{\citenamefont {Karman}\ and\ \citenamefont
  {Hutson}(2018)}]{PhysRevLett.121.2018}%
  \BibitemOpen
  \bibfield  {author} {\bibinfo {author} {\bibfnamefont {T.}~\bibnamefont
  {Karman}}\ and\ \bibinfo {author} {\bibfnamefont {J.~M.}\ \bibnamefont
  {Hutson}},\ }\bibfield  {title} {\bibinfo {title} {Microwave shielding of
  ultracold polar molecules},\ }\href
  {https://doi.org/10.1103/PhysRevLett.121.163401} {\bibfield  {journal}
  {\bibinfo  {journal} {Phys. Rev. Lett.}\ }\textbf {\bibinfo {volume} {121}},\
  \bibinfo {pages} {163401} (\bibinfo {year} {2018})}\BibitemShut {NoStop}%
\bibitem [{\citenamefont {Ni}\ \emph {et~al.}(2010)\citenamefont {Ni},
  \citenamefont {Ospelkaus}, \citenamefont {Wang}, \citenamefont
  {Qu{\'e}m{\'e}ner}, \citenamefont {Neyenhuis}, \citenamefont {De~Miranda},
  \citenamefont {Bohn}, \citenamefont {Ye},\ and\ \citenamefont
  {Jin}}]{ni2010Nature}%
  \BibitemOpen
  \bibfield  {author} {\bibinfo {author} {\bibfnamefont {K.-K.}\ \bibnamefont
  {Ni}}, \bibinfo {author} {\bibfnamefont {S.}~\bibnamefont {Ospelkaus}},
  \bibinfo {author} {\bibfnamefont {D.}~\bibnamefont {Wang}}, \bibinfo {author}
  {\bibfnamefont {G.}~\bibnamefont {Qu{\'e}m{\'e}ner}}, \bibinfo {author}
  {\bibfnamefont {B.}~\bibnamefont {Neyenhuis}}, \bibinfo {author}
  {\bibfnamefont {M.~H.~G.}\ \bibnamefont {De~Miranda}}, \bibinfo {author}
  {\bibfnamefont {J.~L.}\ \bibnamefont {Bohn}}, \bibinfo {author}
  {\bibfnamefont {J.}~\bibnamefont {Ye}},\ and\ \bibinfo {author}
  {\bibfnamefont {D.~S.}\ \bibnamefont {Jin}},\ }\bibfield  {title} {\bibinfo
  {title} {Dipolar collisions of polar molecules in the quantum regime},\
  }\href {https://doi.org/10.1038/nature08953} {\bibfield  {journal} {\bibinfo
  {journal} {Nature}\ }\textbf {\bibinfo {volume} {464}},\ \bibinfo {pages}
  {1324} (\bibinfo {year} {2010})}\BibitemShut {NoStop}%
\end{thebibliography}%

\clearpage

\renewcommand{\figurename}{Fig.}
\renewcommand\thefigure{S\arabic{figure}}
\setcounter{figure}{0}
\renewcommand{\theequation}{S\arabic{equation}}
\setcounter{equation}{0}

\section*{Supplementary Information}

\section*{1. Experimental procedure}

To prepare the $^{23}$Na$^{40}$K molecules in their rovibronic ground state, $2.2 \times 10^{4}$ Feshbach molecules are transferred to the rovibronic ground state $|J = 0, m_J = 0, m_{I_{\mathrm{Na}}} = 3/2, m_{I_{\mathrm{K}}} = -4\rangle$ via a new stimulated Raman adiabatic passage (STIRAP) route. In contrast to previous studies~\cite{park2015Phys.Rev.Lett.,yang2019Science,bause2021Phys.Rev.A} where the intermediate state $B^1\Pi$ $|\nu = 12\rangle \sim c^3\Sigma^{+}$ $|\nu = 35\rangle$ is used, we employ an intermediate state of the spin-orbit coupled complex $B^1\Pi$ $|\nu = 4\rangle \sim c^3\Sigma^{+}$ $|\nu = 25\rangle$, achieving a one-way STIRAP efficiency of approximately $73\%$.  The sample is held in a far-detuned crossed optical trap operating at 1064 $nm$, under an external magnetic field of 72.5 $G$ along $\hat{z}$ direction [see Fig.~\ref{Fig1}(a)]. Details about the preparation process are described in Ref.~\cite{li2023Sci.ChinaPhys.Mech.Astron.}.

Then, the molecules are adiabatically prepared in the dressed state $\ket{+}$ by linearly ramping a dressing microwave field to a specifically chosen intensity within 500 $\mu$s.
Subsequently, the sample is held for 100 $\mu$s to allow stabilization of the microwave dressing power. This is followed by a 500 $\mu$s exposure by the association microwave field with variable detuning, which is used to search for allowed transitions between dressing states and the association of tetratomic molecules. Thereafter, the association field is turned off, and the microwave dressing field is linearly ramped down to zero, adiabatically transferring the remaining molecules from the dressed state $\ket{+}$ back to the initial ground state.

Finally, the molecules are transferred back to the Feshbach molecules via reversed-STIRAP process. Following this, the external magnetic field is ramped from 72.5 $\text{G}$ to 80 $\text{G}$ to cross the Feshbach resonance at 78.3 $\text{G}$  for the dissociation of the Feshbach molecules into a mixture of $^{23}$Na atoms in $|F=1,m_{F}=1\rangle$ and $^{40}$K atoms in $|F=9/2,m_{F}=-9/2\rangle$. The $^{40}$K atoms are then imaged via absorption imaging. We record the number of $^{40}$K to determine the number of molecules initially in the ground state.

\begin{figure}[H]
\centerline{
\includegraphics[width= \linewidth ]{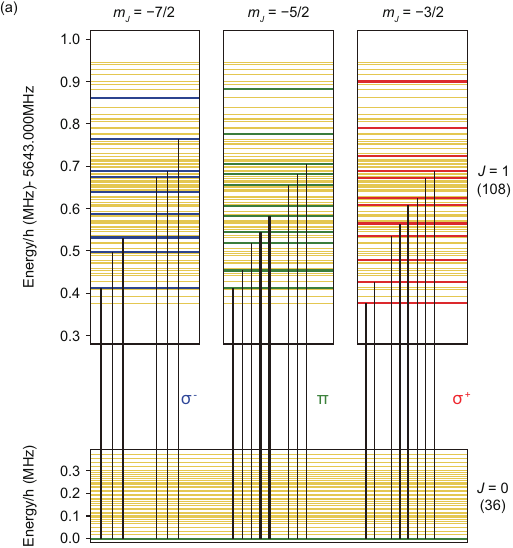}} \vspace{0.1in}
\caption{\textbf{} Energies of 36 hyperfine states of $^{23}$Na$^{40}$K in the rotational ground state $\ket{J=0}$ and 108 states in the first rotational state $\ket{J=1}$ at a external magnetic field of 72 $\text{G}$. The allowed one-photon transition from the absolute ground state $|J=0,m_{J}=0,m_{I_{Na}}=3/2,m_{I_{K}}=-4\rangle$ with $m_{F}=-5/2$ are highlighted with the selection rule $\triangle m_{F}=-1,0,1$ for the microwave filed with $\sigma^{-},\pi,\sigma^{+}$ polarization.
\label{Fig.S1} }
\end{figure}

The hyperfine states of $^{23}$Na$^{40}$K molecules at a magnetic field of 72\,G are shown in Fig.~\ref{Fig.S1}. The microwave frequency of the rotational transition between $\ket{J = 0}$ and $\ket{J = 1}$ is approximately $2B_{\text{rot}} \approx 5.643$\,GHz~\cite{will2016Phys.Rev.Lett.}, where $B_{\text{rot}}$ is the rotational constant. There are 36 hyperfine states in the ground rotational level $\ket{J = 0}$ and 108 states in the first excited rotational level $\ket{J = 1}$.

At a magnetic field of 72\,G, hyperfine interactions strongly couple the rotational and nuclear spin degrees of freedom. The only good quantum number is the projection of the total angular momentum, defined as $m_{F} = m_{J} + m_{I_{\text{Na}}} + m_{I_{\text{K}}}$, where $m_{J}$ is the projection of the rotational angular momentum, and $m_{I_{\text{Na}}}$ and $m_{I_{\text{K}}}$ are the projections of the nuclear spins of $^{23}$Na and $^{40}$K, respectively, onto the quantization axis.
Molecules in the lowest hyperfine state $\ket{J = 0, m_{J} = 0, m_{I_{\text{Na}}} = 3/2, m_{I_{\text{K}}} = -4}$ with $m_{F} = -5/2$ can be coupled to the first rotational level with $m_{F} = -7/2$, $-5/2$, or $-3/2$ by selecting appropriate polarizations of the microwave field. For each allowed one-photon transition, the lines are highlighted and bolded to indicate those with larger transition dipole moments. The state energies and transition strengths are calculated using the Python-based program described in Ref.~\cite{blackmore2023ComputerPhysicsCommunications}.

\section*{2. Dressed states picture}

The Hamiltonian for a multi-level system consisting of a single ground state
\(|a\rangle\) with zero energy and \(N\) excited states \(|b_j\rangle\)
with energies \(\omega_{b_j}\), interacting with a microwave field of frequency
\(\omega_d\), is given by

\begin{equation}
\begin{aligned}
H ={} & \hbar \, \omega_d \, c^\dagger c + \sum_{j=1}^{N} \hbar \, \omega_{b_j} \, |b_j\rangle\langle b_j| \\
       & + \sum_{j=1}^{N} \hbar \, g_j (c^\dagger + c)\Big( |a\rangle\langle b_j|
                                      + \, |b_j\rangle\langle a| \Big).
\end{aligned}
\end{equation}
Under the rotating wave approximation (RWA), the Hamiltonian is simplified as
\begin{equation}
\begin{aligned}
H_{\mathrm{RWA}} ={} & \hbar \, \omega_d \, c^\dagger c + \sum_{j=1}^{N} \hbar \, \omega_{b_j} \, |b_j\rangle\langle b_j| \\
       & + \sum_{j=1}^{N} \hbar \, g_j \Big(c^\dagger |a\rangle\langle b_j|
                                      + c\, |b_j\rangle\langle a| \Big).
\end{aligned}
\end{equation}
Here, the total excitation number $\hat{N} = c^\dagger c + \sum_{j} |b_j\rangle \langle b_j|$ is conserved with $[\hat{N}, H_{\mathrm{RWA}}] = 0$.
In a subspace with a fixed total excitation number, the Hamiltonian can be expressed as
\begin{align*}
H_{s}= \hbar\Delta\vert a \rangle \langle a \vert + \sum_{j = 1}^{N} (\varepsilon_j \vert b_j \rangle \langle b_j \vert + \frac{\hbar\Omega_j}{2} \vert b_j \rangle \langle a \vert + \text{H.c})
\end{align*}
where $\Delta=\omega_{d}-\omega_{0}$ is the detuning of microwave dressing field, $\Omega_j=2g_{j}\sqrt{n+1}$ is the Rabi coupling strength between the ground state and the excited state $| b_j \rangle$, and $n$ denotes the microwave photon number.

We can diagonalize the Hamiltonian to obtain its eigenvalues and eigenstates. In the regime where $\Delta,\, \Omega_j \gg \varepsilon_j$, the eigenenergies of the two dressed states can be approximated as
\begin{equation}
E_{\pm} \approx \frac{\hbar\Delta}{2} \pm \frac{\hbar}{2} \sqrt{\Delta^2 + \Omega_1^2 + \Omega_2^2 + \cdots + \Omega_N^2},
\end{equation}
while the remaining $N - 1$ dark states have zero eigenenergy and are decoupled from the ground state.

Following, we consider a simplified model of a three levels system (one ground state $| a; n \rangle$, and two hyperfine states in first rotational state $| b_1; n-1 \rangle$, $| b_2; n-1 \rangle$), which can be described as
\begin{align*}
 H_{\mathrm{eff}}=\hbar\left(\begin{array}{lll}
\Delta & \frac{\Omega_{1}}{2} & \frac{\Omega_{2}}{2} \\
\frac{\Omega_{1}}{2} & 0 & 0 \\
\frac{\Omega_{2}}{2} & 0 & 0
\end{array}\right).
\end{align*}
The corresponding eigenenergies are
\begin{align*}
E_{d} & = 0, \\
E_{+} & = \frac{\hbar\Delta}{2}+\frac{\hbar}{2} \sqrt{\Delta^{2}+\Omega_{1}^{2}+\Omega_{2}^{2}}, \\
E_{-} & = \frac{\hbar\Delta}{2}-\frac{\hbar}{2} \sqrt{\Delta^{2}+\Omega_{1}^{2}+\Omega_{2}^{2}}.
\end{align*}
The energy splitting between the two dressed states is given by $\hbar\Omega_{\text{eff}}=\hbar\sqrt{\Delta^{2}+\Omega_{1}^{2}+\Omega_{2}^{2}}$. The corresponding dressed eigenstates can be expressed as
\begin{align*}
\ket{0}_n &= -\sin\phi \ket{b_1;n-1} + \cos\phi \ket{b_2;n-1}, \\
\ket{+}_n &= \cos\theta \ket{a,n} \\
&+ \sin\theta \cos\phi \ket{b_1;n-1} + \sin\theta \sin\phi \ket{b_2;n-1}, \\
\ket{-}_n &= -\sin\theta \ket{a,n} \\
&+ \cos\theta \cos\phi \ket{b_1;n-1} + \cos\theta \sin\phi \ket{b_2;n-1}, \\
\end{align*}
where the mixing angles are defined by
\begin{align*}
\tan\phi &=\frac{\Omega_{2}}{\Omega_{1}},\\
\tan 2\theta &=\frac{\sqrt{\Omega^2_{1}+\Omega^2_{2}}}{\Delta}=\frac{\sqrt{\Omega^2_{\text{eff}}-\Delta^2}}{\Delta},
\end{align*}
with $0\leq\phi\leq\pi/4$ and $0\leq\theta<\pi/2$.

When a blue-detuned microwave dressing field is applied, a second microwave field is used to drive the allowed transitions within the dressing state system, the transition dipole moments as a function of the mixing angles are expressed as
\begin{align}
_{n+1} \langle -|\hat{d}|+ \rangle_n &= \cos^2\theta(\cos\phi \ \textbf{d}_{ab_{1}}+\sin\phi \ \textbf{d}_{ab_{2}}),  \\
_{n+1} \langle 0|\hat{d}|+ \rangle_n &= \cos\theta (-\sin\phi \ \textbf{d}_{ab_{1}}+\cos\phi \ \textbf{d}_{ab_{2}}),  \\
_{n-1} \langle 0|\hat{d}|+ \rangle_n &= 0, \\
_{n-1} \langle -|\hat{d}|+ \rangle_n &= -\sin^2\theta(\cos\phi \ \textbf{d}_{ab_{1}}+\sin\phi \ \textbf{d}_{ab_{2}}),
\end{align}
where $\hat{d}$ is the dipole operator for the dressed states, $\textbf{d}_{ab_{1}}=\langle a|\hat{d}|b_{1}\rangle$ and $\textbf{d}_{ab_{2}}=\langle a|\hat{d}|b_{2}\rangle$ are the transition dipole moments for the bare states.

The vanishing dipole moment for the $\ket{+}_{n} \leftrightarrow \ket{0}_{n-1}$ transition can be attributed to the conservation of the total excitation number in the dressing-state model described above. We notice that the microwave source is an infinite reservoir of microwave photon in our experiment, the molecular system can exchange an arbitrary number of photon with the field. This results in the manifold $\{| a; n \rangle, | b_1; n-1 \rangle, | b_2; n-1 \rangle \}$ is not isolated, the total excitation number is no longer conserved, the block-diagonal structure of the Hamiltonian is broken. Then the dipole moment of $\ket{+}_{n}$ $\leftrightarrow$ $\ket{0}_{n-1}$ is not zero.

For a microwave dressing field with blue detuning, the mixing angle is $0\leq \theta<\pi/4$. In this regime, the ratio of the transition dipole moments for the transitions $\ket{+}_{n}$ $\leftrightarrow$ $\ket{-}_{n+1}$ and $\ket{+}_{n+1}$ $\leftrightarrow$ $\ket{-}_{n}$ is given by $\cos^{2}(\theta)/\sin^{2}(\theta)>1$. Meanwhile, the transition dipole moment of $\ket{+}_{n}$ $\leftrightarrow$ $\ket{0}_{n+1}$ is significantly larger than that of $\ket{+}_{n}$ $\leftrightarrow$ $\ket{0}_{n-1}$. As a result, the dressed-state spectroscopy exhibits a more pronounced loss feature at the negative value $\Delta_{A}$.

\begin{figure}[t]
\centerline{
\includegraphics[width= 3 in ]{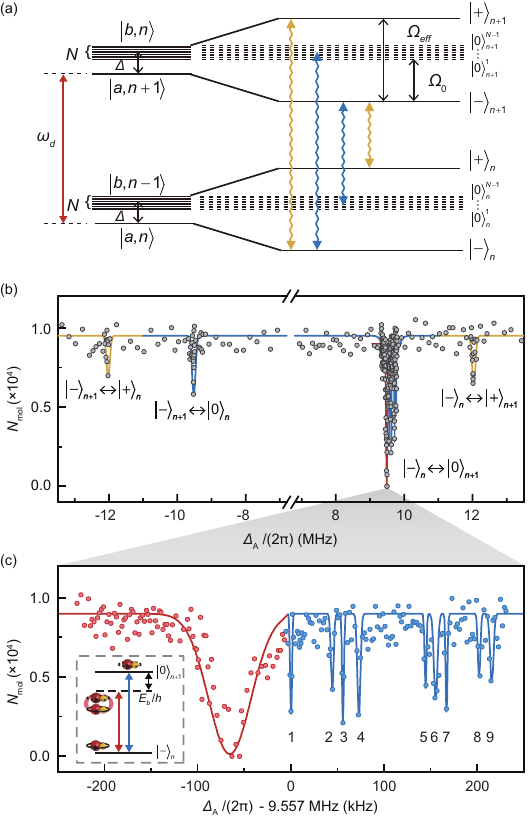}} \vspace{0.01in}
\caption{\textbf{} (a) Illustration of the allowed transitions in the dressed molecules system under a red-detuned dressing microwave field. The left side depicts the uncoupled basis states, consisting of the ground state $\ket{a,n}$ and $N$ hyperfine excited states $\ket{b,n-1}$. The right side shows the corresponding dressed-state picture. Wavy lines represent the four allowed transitions: yellow lines indicate transitions between the two dressed states, and blue lines represent transitions between the dressed states and the dark states.
(b) Loss spectrum of the molecules initially prepared in the dressed state $\ket{-}$ as a function of the microwave difference $\Delta_A = \omega_a - \omega_d$. The loss peaks indicate the resonant transition frequencies corresponding to the four allowed transitions from the dressed state $\ket{-}$. The microwave dressing field is applied with a coupling strength of $\Omega = 2\pi \times 9.913\,\mathrm{MHz}$ and a red detuning of $\Delta = 2\pi \times -7\,\mathrm{MHz}$.
(c) Zoomed-in structure of the spectrum around the resonant transition $\ket{-}_{n}$ $\leftrightarrow$ $\ket{0}_{n+1}$. Besides the nine loss peaks marked by blue lines, which originate from transitions between dressed molecular states, a new loss peak marked by a red line signifies the association of tetratomic molecules.
\label{Fig.S2} }
\end{figure}

For a microwave dressing field with red detuning, the molecules are prepared in the dressed state $\ket{-}$. Fig.~\ref{Fig.S2}(a) illustrates the allowed transitions under such a red-detuned dressing field. As discussed above, the dressed-state spectroscopy exhibits a more pronounced loss feature at the positive microwave difference $\Delta_{A}$. We focus on the detailed loss spectrum of the $\ket{-}_{n} \leftrightarrow \ket{0}_{n+1}$ transition, as shown in Fig.~\ref{Fig.S2}(c). An additional loss peak also is observed near the first free--free dressed-molecule transition $\ket{-}_{n} \rightarrow \ket{0}^{1}_{n+1}$, occurring at a frequency about $100~\text{kHz}$ lower than the first transition line. This feature can be attributed to the formation of tetratomic molecules via microwave association.

\section*{3. Tetramer states}

\begin{figure}[ptb]
\centering
\includegraphics[width=\linewidth]{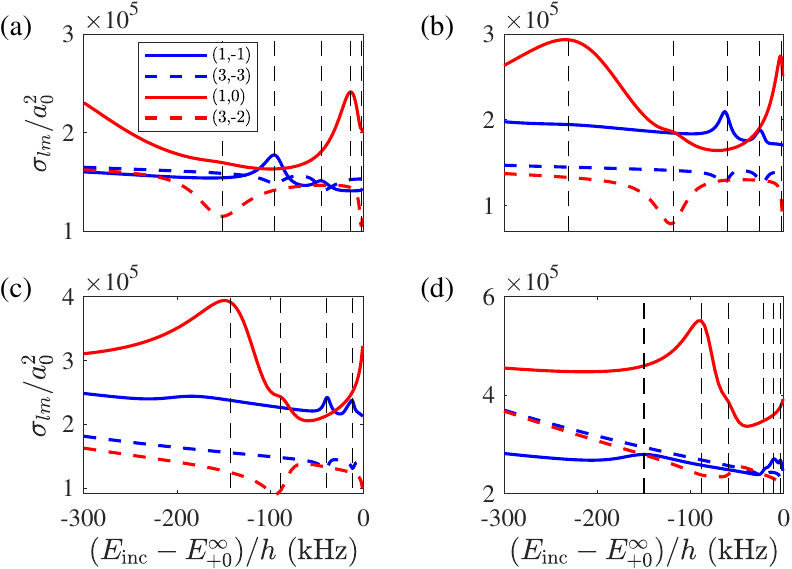}
\caption{Scattering cross sections across resonances with the tetramer states, where Rabi frequencies are (a) $\Omega=2\protect\pi\times 10$~MHz,
(b) $\Omega=2\protect\pi\times 9$~MHz, (c) $\Omega=2\protect\pi\times 8$%
~MHz, and (d) $\Omega=2\protect\pi\times 7$~MHz. Other parameters are $%
\Delta=2\protect\pi\times 7$~MHz and the ellipticity of
the dressing microwave $\protect\xi=5^\circ$. Here, $\protect%
\sigma_{lm}$ represents the scattering cross section from incident partial
wave channel $(l,m)$ to outgoing partial wave channel $(l,m)$ in the dressed
state $|+-\rangle_s$, and dark dashed vertical lines denote the center
positions of the resonances.}
\label{fig_resonance}
\end{figure}

To investigate the properties of tetramer states, we perform multi-channel scattering calculations following the approach in previous studies~\cite{chen2024Nature,deng2024}. The collision between two molecules is governed by the
dipolar interaction
\begin{equation}
V(\mathbf{r})=\frac{d^{2}}{4\pi \epsilon _{0}r^{3}}\left[ \hat{\mathbf{d}}%
_{1}\cdot \hat{\mathbf{d}}_{2}-3(\hat{\mathbf{d}}_{1}\cdot \hat{\mathbf{r}})(%
\hat{\mathbf{d}}_{2}\cdot \hat{\mathbf{r}})\right] ,  \label{DDI}
\end{equation}
where $\epsilon _{0}$ is the vacuum permittivity, $r=|\mathbf{r}|$ and $\hat{\mathbf{r}}=\mathbf{r}/r$ are determined by the relative distance $\mathbf{r}$, $d$ is the electric dipole moment of NaK molecules, and $\hat{\mathbf{d}}_{j=1,2}$ denotes the unit vector along the internuclear axis of the $j$th molecule. We consider tetramer bound states in the symmetric dressed-state channels of the two-molecule Hilbert space. Unlike previous studies, we focus on the tetramers in the $|+0\rangle _{\mathrm{s}}$ channel, where the symmetric states are defined as $|ij\rangle _{\mathrm{s}}=(|ij\rangle+|ji\rangle )/\sqrt{2}$.

To determine the binding energy $\mathcal{E}_{b}$ and linewidth $\gamma _{b}$ of the tetramer states, we study collisions with incident energy $E_{\mathrm{inc}}$ below $E_{+0}^{\infty }$ ($E_{\alpha }^{\infty }$ is the asymptotic energy of the corresponding state $|\alpha \rangle _{\mathrm{s}}$). As shown in Fig.~\ref{Fig1}(d) in the main text, tuning the incident energy $E_{\mathrm{inc}}$ induces Feshbach resonances when $E_{\mathrm{inc}}$ matches the binding energy of a tetramer state, where channels with $E_{\alpha }^{\infty }>E_{\mathrm{inc}}$ act as closed channels and the others as open channels. The resonance peak position and width then yield $\mathcal{E}_{b}$ and $\gamma _{b}$, respectively. To compute the scattering cross section, we perform coupled-channel calculations including all dressed-state channels. For quantitative accuracy, we add a van der Waals potential $-C_{\mathrm{vdW}}/r^{6}$ in $V(\mathbf{r})$ to account for universal background scattering~\cite{PhysRevLett.104.2010}, and incorporate an absorption boundary condition inside the shielding core to model losses from four-body complex formation~\cite{PhysRevLett.121.2018}.

\begin{figure}[t]
\centerline{
\includegraphics[width= \columnwidth]{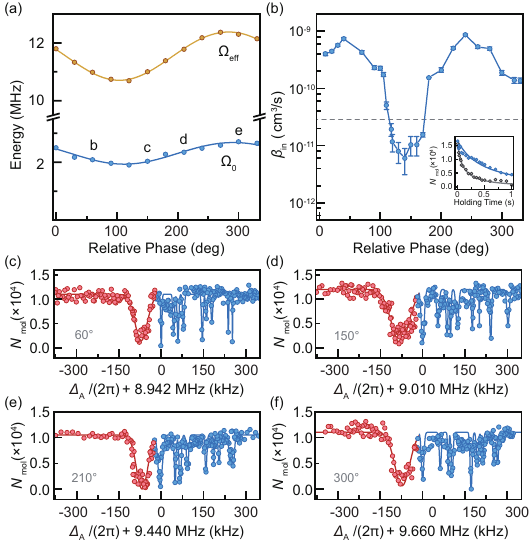}} \vspace{0.01in}
\caption{\textbf{}
(a) Energy splittings $\Omega_{\text{eff}}$ between the two dressed states and $\Omega_{0}$ between the dark states $\ket{0}^{1}_{n}$ and the dressed state $\ket{+}$, measured at blue detuning $\Delta = 2\pi \times 7\,\mathrm{MHz}$, shown as functions of the relative phase between the two paths of the doubly fed waveguide.
(b) Extracted two-body loss rate as a function of the relative phase. The horizontal dashed line denotes inelastic two-body loss rate coefficient in the unshielded case. The inset displays the molecules number versus holding time in the optical trap with and without the microwave shielding.
(c--f) Enlarged views of the spectra around the resonant transition $\ket{+}_{n} \leftrightarrow \ket{0}_{n+1}$. In addition to the nine loss peaks (blue lines) arising from transitions between dressed molecular states, an additional loss peak (red line) is observed, indicating the association of tetratomic molecules.
\label{Fig.S4}}
\end{figure}

As shown in Fig.~\ref{fig_resonance}, resonances appear in different angular-momentum channels $(l,m)$ of the dressed state $|+0\rangle _{\mathrm{s}}$, corresponding to binding energies of tens to hundreds of kHz. The presence of multiple resonances indicates that the potential supports several tetramer bound states. Owing to the strong dipolar interaction, many tetramer states exist in different angular-momentum channels, giving rise to both $p$-wave and high-partial-wave resonances. Since the $|+0\rangle_{s}$ channel is not well shielded, all these states are short-lived. Moreover, the deeper the tetramer state, the shorter its lifetime, exhibiting linewidths on the order of tens to hundreds of kHz. These states have similar binding energies and large linewidths, rendering them experimentally indistinguishable. As the Rabi frequency decreases, the dipolar interaction weakens, causing all the binding energies to shift toward the threshold.

\
\section*{4. The dressed state spectroscopy under microwave dressing fields with different polarization}

We also measured the dressed state spectroscopy under microwave dressing fields with different polarizations, which were controlled by adjusting the relative phase between two orthogonal polarizations in the doubly-fed waveguide. The energy splittings $\Omega_{\text{eff}}$ between two dressed states and $\Omega_{0}$ between the dressed state $\ket{-}_{n}$ and the dark state $\ket{0}^{1}_{n}$, as functions of the relative phase, are shown in Fig.~\ref{Fig.S4}(a).
Importantly, the loss structure associated with the tetramer state and the transition $\ket{+}_{n}$ $\leftrightarrow$ $\ket{0}_{n+1}$ remains nearly unchanged across different polarizations (see Fig.~\ref{Fig.S4} (c,d,e,f)). This observation indicates that the microwave association of tetratomic molecules in our approach exhibits robustness and insensitivity to the polarization of the microwave dressing field.

When the relative phase was turned to $140^\circ$, we observed microwave shielding of $^{23}$Na$^{40}$K molecules with lowest  two-body loss rate, as shown in Fig.~\ref{Fig.S4} (b). We extracted two-body loss rate from the time evolution of the molecules number versus holding time in the optical trap using the rate equation~\cite{ni2010Nature}.
\begin{align*}
\frac{d n}{d t}=-\beta_{in} n^{2}-\Gamma n.
\end{align*}
with inelastic two-body loss rate coefficient $\beta_{in}$ and one-body loss rate $\Gamma$. The extracted two-body loss rate decreases significantly from 2.85$\times10^{-11}$ cm$^{3}$s$^{-1}$ in the unshielded case to 5.98$\times10^{-12}$ cm$^{3}$s$^{-1}$, about one order of magnitude smaller.

\end{document}